\documentclass[sn-mathphys,Numbered]{sn-jnl}% Math and Physical Sciences Reference Style
\usepackage{graphicx}%
\usepackage{multirow}%
\usepackage{amsmath,amssymb,amsfonts}%
\usepackage{amsthm}%
\usepackage{mathrsfs}%
\usepackage[title]{appendix}%
\usepackage{xcolor}%
\usepackage{textcomp}%
\usepackage{manyfoot}%
\usepackage{booktabs}%
\usepackage{algorithm}%
\usepackage{algorithmicx}%
\usepackage{algpseudocode}%
\usepackage{listings}%
\usepackage{url}%
\usepackage[all]{xy}%

\usepackage{bussproofs}%
\usepackage{scalerel}[2014/03/10]%
\usepackage{tikz}%

\definecolor{dkgreen}{rgb}{0,0.6,0}
\definecolor{ltblue}{rgb}{0,0.4,0.4}
\definecolor{dkblue}{rgb}{0,0.2,0.8}
\definecolor{dkviolet}{rgb}{0.3,0,0.5}

\lstdefinelanguage{Coq}{ 
    mathescape=true,
    texcl=false, 
    morekeywords=[1]{Section, Module, End, Require, Import, Export,
        Variable, Variables, Parameter, Parameters, Axiom, Hypothesis,
        Hypotheses, Notation, Local, Tactic, Reserved, Scope, Open, Close,
        Bind, Delimit, Definition, Let, Ltac, Fixpoint, CoFixpoint, Add,
        Morphism, Relation, Implicit, Arguments, Unset, Contextual,
        Strict, Prenex, Implicits, Inductive, CoInductive, Record,
        Structure, Canonical, Coercion, Context, Class, Global, Instance,
        Program, Infix, Theorem, Lemma, Corollary, Proposition, Fact,
        Remark, Example, Proof, Goal, Save, Qed, Defined, Hint, Resolve,
        Rewrite, View, Search, Show, Print, Printing, All, Eval, Check,
        Projections, inside, outside, Def},
    morekeywords=[2]{forall, exists, exists2, fun, fix, cofix, struct,
        match, with, end, as, in, return, let, if, is, then, else, for, of,
        nosimpl, when},
    morekeywords=[3]{Type, Prop, Set, true, false, option},
    morekeywords=[4]{pose, set, move, case, elim, apply, clear, hnf,
        intro, intros, generalize, rename, pattern, after, destruct,
        induction, using, refine, inversion, injection, rewrite, congr,
        unlock, compute, ring, field, fourier, replace, fold, unfold,
        change, cutrewrite, simpl, have, suff, wlog, suffices, without,
        loss, nat_norm, assert, cut, trivial, revert, bool_congr, nat_congr,
        symmetry, transitivity, auto, split, left, right, autorewrite},
    morekeywords=[5]{by, done, exact, reflexivity, tauto, romega, omega,
        assumption, solve, contradiction, discriminate},
    morekeywords=[6]{do, last, first, try, idtac, repeat},
    morecomment=[s]{(*}{*)},
    showstringspaces=false,
    morestring=[b]",
    morestring=[d],
    tabsize=3,
    extendedchars=false,
    sensitive=true,
    breaklines=false,
    basicstyle=\small,
    captionpos=b,
    columns=[l]flexible,
    identifierstyle={\ttfamily\color{black}},
    keywordstyle=[1]{\ttfamily\color{dkviolet}},
    keywordstyle=[2]{\ttfamily\color{dkgreen}},
    keywordstyle=[3]{\ttfamily\color{ltblue}},
    keywordstyle=[4]{\ttfamily\color{dkblue}},
    keywordstyle=[5]{\ttfamily\color{dkred}},
    stringstyle=\ttfamily,
    commentstyle={\ttfamily\color{dkgreen}},
    literate=
    {==}{{\code{==}\;}}1
    {==>}{{\code{==>}\;}}1
    {\/\\}{{$\wedge\;$}}1
    {\\\/}{{$\vee\;$}}1
    {'}{$^\prime$}1
}[keywords,comments,strings]

\theoremstyle{thmstyleone}%
\newtheorem{theorem}{Theorem}%  meant for continuous numbers
\newtheorem{corollary}[theorem]{Corollary}% 
\newtheorem{lemma}[theorem]{Lemma}% 

\theoremstyle{thmstyletwo}%
\newtheorem{example}{Example}%

\theoremstyle{thmstylethree}%
\newtheorem{definition}{Definition}%

\begin{document}

\title[Article Title]{DRAFT: A Formally Verified Constructive Proof of the Consistency of Peano Arithmetic Using Ordinal Assignments}


\author[1]{\fnm{Aaron} \sur{Bryce}}\email{Aaron.Bryce@anu.edu.au}

\author*[2]{\fnm{Rajeev} \sur{Gor\'e}}\email{rajeev.gore@monash.edu}

\affil[1]{\orgdiv{Mathematical Sciences Institute}, \orgname{Australian National University}, \orgaddress{\city{Canberra}, \postcode{2600}, \state{ACT}, \country{Australia}}}

\affil*[2]{\orgdiv{Faculty of Information Technology}, \orgname{Monash University}, \orgaddress{\street{Street}, \city{Clayton}, \postcode{3800}, \state{Victoria}, \country{Australia}}}


\abstract{Gentzen's 1936 proof of the consistency of Peano Arithmetic was a significant result in the foundations of
  mathematics. We provide here a modified version of the proof, based on G\"{o}del's reformulation, and including additional
  details and minor corrections which are necessary to definitively prove the well-foundedness of the cut-elimination argument in
  a constructive environment. All results have been verified using the Coq theorem prover.

  NOTE TO READERS 26 February 2026: this is a draft which we had intended to submit to the Journal of Automated Reasoning with no particular time-line in our minds as the work was completed as part of Aaron's honours project at ANU in 2023. For that reason, we have used the Springer style files. We are putting it on arxiv as there appears to be some interest in this work as indicated by a post to
  \url{https://proofassistants.stackexchange.com/questions/6462/how-far-is-gentzens-consistency-proof-of-peano-arithmetic-from-being-formalized}
  in early February 2026. The Coq code is available here: \url{https://github.com/aarondroidbryce/Gentzen/tree/master }}




\maketitle

\newpage{}
\tableofcontents{}
\newpage{}

\section{Introduction}

\subsection{A Crisis of Foundations}

During the 1930's mathematics was progressing rapidly, however not all of these developments were positive. Hilbert's Program, a series of results set forth at the turn of the 20th century that would have solidified mathematics down to an unshakeable logical foundation, suffered it first major defeat. G\"{o}del proved his two famous incompleteness theorems, and Hilbert's Program as it was originally envisioned was doomed\cite{zach}. Ideas of proving all of mathematics consistent seemed to be doomed if something as simple as the consistency of arithmetic was unable to be proven. Many took G\"{o}del's results as the end and moved on to other pursuits, however not all gave up hope choosing to reject the impossibility reading\cite{hilbert}. Gerhard Gentzen continued to work on his proof, his previous work on sequent calculi\cite{gentzen-uber} giving him unique insights which allowed him to work around the constraints placed by the incompleteness theorem. His initial proof\cite{gentzen-other} was considered unconvincing, but after several of his own revisions and another by G\"{o}del\cite{godel}, his work stands as the first finitary\footnote{Some academics debate if this proof truly counts as finitary, because Hilbert never specified what he meant by a proof by finitary means, see\cite{chow}.} consistency proof for arithmetic.

\subsection{Incompleteness and Consistency}

For notational convenience the statement ``arithmetic is consistent'' is represented as $Cons_{PA}$, where $PA$ is referring to the Peano Arithmetic, which is one of the most common axiomatisations of arithmetic\cite{fritz}. The main difficulty in proving this result is G\"{o}del's Second Incompleteness Theorem and its corollaries.

\begin{theorem}[G\"{o}del's Second Incompleteness Theorem]
\label{theorem:Incomplete}
For any Hilbert calculus $L$, if $L$ contains Robinson Arithmetic and is consistent, then $L$ can not prove $Cons_{L}$\cite{godel}.
\end{theorem}

\begin{corollary}
\label{corollary:notself}
Peano Arithmetic can not prove its own consistency.
\end{corollary}

Furthermore, because Peano Arithmetic can prove the consistency of Primitive Recursive Arithmetic we have an additional corollary due to transitivity of consistency.

\begin{corollary}
\label{corollary:notPRA}
Primitive Recursive Arithmetic can not prove the consistency of Peano Arithmetic.
\end{corollary}

As such, it was necessary for Gentzen to find an alternative formulation $L$, such that Peano Arithmetic could not prove the consistency of $L$, but $L$ could prove the consistency of PA. To this end Gentzen created the Hilbert calculus PRA + $\epsilon_0$, which was independent of PA since it contained as an axiom of transfinite induction up to the largest finitely representable ordinal, $\epsilon_0$, something which was not provable within PA. Then Gentzen used PRA + $\epsilon_0$ to prove the consistency of the sequent calculus PA$_\omega$, as well as the equivalence of the consistency of PA$_\omega$ and PA, thereby arriving at a consistency proof of PA which did not violate G\"{o}del's Incompleteness.

\begin{example}[Relationship between the different logics in Gentzens proof]
\
\xymatrixcolsep{5pc}
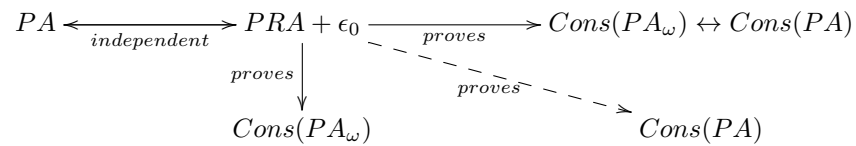
\xymatrix{
    {PA}
    \ar[r]_-{independent}
    & {PRA + \epsilon_0}
    \ar[l]
    \ar[r]_-{proves}
    \ar[d]_-{proves}
    \ar@{-->}[rd]_-{proves}
    & {Cons(PA_\omega)  \leftrightarrow Cons(PA)}
    \\
    {} & {Cons(PA_\omega)} & {Cons(PA)}
  }
\end{example}

\subsection{Finiteness and Formalisation}

Gentzen's method for proving consistency used the now familiar technique of Cut Elimination over a sequent calculus, a method which he had pioneered \cite{gentzen-other}. He constructed the logic PA$_\omega$, which used 7 consistency preserving deductive rules and the Cut rule, which does not preserve consistency. The argument then follows by showing that any derivation in PA$_\omega$ which uses the Cut rule can be transformed into a different derivation, which does not rely on the Cut rule. Finally, Gentzen shows that PA$_\omega$ is equivalent to the normal construction of arithmetic from the Peano axioms.

A critical part of Gentzen's argument comes from the assigning of ordinal numbers to the proof trees of PA$_\omega$. It is using this assignment and the well-foundedness of the ordinals through which he is able to argue for the termination of the proposed Cut Elimination algorithm. However, while the ordinal calculations for the Cut Elimination argument are mostly understood, they are usually not present in enough detail to make a constructive argument. For example, both ``Introduction to Mathematical Logic''\cite{mendelson} and ``Basic Proof Theory''\cite{proof-theory} provide versions of the argument which do not include the explicit heights of the derivations used in proving the equivalence of the logical calculi PA$_\omega$ and PA. Concerningly, this oversight actually draws into question the validity of the overall proof, since it means that there is no guarantee that the representations of the Peano Axioms in PA$_\omega$ are assigned ordinals which fall within the bounds required for Cut Elimination. By calculating these assignments it is also possible to gain a deeper understanding of the proof, since it shows that Gentzen's approach could not be performed using a weaker logic than that of Primitive Recursive Arithmetic strengthened with transfinite induction up to $\epsilon_0$ (PRA + $\epsilon_0$). Additionally, as is mentioned in Timothy Chow's Paper on the Consistency of Arithmetic\cite{chow}, there is still some argument about the value/reasonableness of Gentzen's proof. As such, building a fully constructive version of the argument within Coq helps provide more solidity. Of course, as mentioned by Chow, at some point you must accept some set of axioms, and this work does rely upon the axioms of the Calculus of Inductive Constructions being sound. However, since Gentzen's proof only relies on the axioms of PRA+$\epsilon_0$, we should not actually be relying on most of the axioms available to us, and importantly, due to its constructive nature, things such as the finite approximations to consistency could feasibly be calculated using the algorithms we developed.

For these reasons we created a computer verified version of Gentzen's proof within Coq, which would be fully constructive and help close this gap in the literature surrounding such an important part of mathematical logic. In doing so we uncovered several generalisations which are often used in discussions of the proof which are either incorrect or specific to a certain construction of the underlying calculi, and produced separate arguments for such cases as necessary. We discuss these is Sections \ref{sec:LEM}, \ref{sec:PA_equiv} and \ref{sec:godel}.

\section{Components of Gentzen's Proof}

Before getting into the cut elimination argument of PA$_\omega$, we will need to construct the various components set out above, to build a framework to prove the result within. First we must construct the Hilbert calculus for Peano arithmetic. Then we provide a construction for the ordinals less than $\epsilon_0$ as they will occur throughout the entire argument. We must also define the sequent calculus PA$_\omega$, and the notion of a derivation within the calculus. Finally we need to relate the sequent calculus to the Hilbert calculus for Peano arithmetic. With that in place we will be able to move on to the cut elimination argument in full.

\

\subsection{Hilbert Calculus for Peano Arithmetic}
The first component needed is a construction of Peano Arithmetic as a Hilbert calculus. This will be done with an inductive structure similar to that which we will use for PA$_\omega$, each axiom and deductive rule will be given an associated constructor with the correct arity. The necessary aspects of Peano arithmetic can be broken up into 4 classes, the First Order Logic Axioms, the Arithmetic Axioms, the Deductive Rules and the axiom schema for Induction.

\begin{definition}[First-order logic axioms]
The first order axioms define the relationship between implication and the logical connectives of implication, negation and quantification.

\begin{description}
\item[FOL1:] $\forall \ A \ B, \ \ A \to (B \to A)$
\item[FOL2:] $\forall \ A \ B \ C, \ \ (A \to (B \to C)) \to ((A \to B) \to (A \to C))$
\item[FOL3:] $\forall \ A \ B, \ \ (\neg A \to \neg B) \to ((\neg A \to B) \to A)$
\item[FOL4:] $\forall \ A[x_i] \ \ y, \ \ (\forall m, A[x_i/m]) \to A[x_i/y]$
\item[FOL5:] $\forall \; B \; A[x_i], \; ((\forall m, (B \to P)[x_i/m]) \to (B \to \forall m, A[x_i/m])$ if $x_i$ does not appear in $B$.   
\end{description}
\end{definition}

\begin{lstlisting}[language=Coq,label=list:PeanoCoq,abovecaptionskip=-\medskipamount]
Inductive Peano_Theorem : formula -> Type :=
| FOL1 : forall (A B : formula),
          Peano_Theorem (A -> (B -> A))
| FOL2 : forall (A B C : formula),
          Peano_Theorem ((A -> (B -> C)) -> ((A -> B) -> (A -> C)))
| FOL3 : forall (A B : formula),
          Peano_Theorem ((neg A -> neg B) -> ((neg A -> B) -> A))
| FOL4 : forall (A : formula) (n : nat) (t : term),
          closed_t t = true ->
            Peano_Theorem (lor (neg(univ n A)) (substitution A n t))
| FOL5 : forall (A B : formula) (n : nat),
          member n (free_list A) = false ->
            Peano_Theorem ((univ n (A -> B)) -> (A -> (univ n B)))
...
\end{lstlisting}

\begin{definition}[Arithmetic axioms]The arithmetic axioms contain the relationships between equality, successor, addition and multiplication for bound variables.
\begin{description}
\item{Eq\_trans:} $\forall \ m \ n \ p, \ \ ([x_i/m] = [x_j/n]) \to (([x_j/n] = [x_k/p]) \to [x_i/m] = [x_k/p])$
\item{Eq\_succ :} $\forall \ m \ n, \ \ ([x_i/m] = [x_j/n]) \to (S [x_i/m] = S [x_j/n])$
\item{Non\_zero :} $\forall \ m, \ \ \neg (S [x_i/m] = 0)$ 
\item{Succ\_eq :} $\forall \ m \ n, \ \ (S [x_i/m] = S [x_j/n]) \to ([x_i/m] = [x_j/n])$
\item{Pl\_zero :} $\forall \ m, \ \ ([x_i/m] + 0 = [x_i/m])$
\item{Pl\_succ :} $\forall \ m \ n, \ \ ([x_i/m] + S [x_j/n] = S ([x_i/m] + [x_j/n]))$
\item{Ml\_zero :} $\forall \ m, \ \ ([x_i/m] \times 0 = 0)$
\item{Ml\_succ :} $\forall \ m \ n, \ \ ([x_i/m] \times S [x_j/n] = ([x_i/m] \times [x_j/n]) + [x_i/m])$\
\end{description}
\end{definition}

\begin{lstlisting}[language=Coq,label=list:PeanoCoq,abovecaptionskip=-\medskipamount]
Inductive Peano_Theorem : formula -> Type :=
...
| equ_trans : Peano_Theorem
                (univ 0
                  (univ 1
                    (univ 2
                      ((atom (equ (var 0) (var 1))) ->
                        ((atom (equ (var 1) (var 2))) ->
                            (atom (equ (var 0) (var 2)))))))) 
| equ_succ : Peano_Theorem
                (univ 0
                  (univ 1
                    ((atom (equ (var 0) (var 1))) ->
                      (atom (equ (succ (var 0)) (succ (var 1)))))))
| non_zero : Peano_Theorem (univ 0 (neg (atom (equ zero (succ (var 0))))))
| succ_equ : Peano_Theorem
              (univ 0
                (univ 1
                  ((atom (equ (succ (var 0) succ (var 1)))) ->
                    (atom (equ (var 0) (var 1))))))
| pl0 : Peano_Theorem (univ 0 (atom (equ (plus (var 0) zero) (var 0))))
| plS : Peano_Theorem
          (univ 0
            (univ 1
              (atom (equ (plus (var 0) (succ (var 1))) (succ (plus (var 0) (var 1)))))))
| ml0 : Peano_Theorem (univ 0 (atom (equ (times (var 0) zero) zero)))
| mlS : Peano_Theorem
          (univ 0
            (univ 1
                (atom (equ (times (var 0) (succ (var 1)))
                            (plus (times (var 0) (var 1)) (var 0))))))
...
\end{lstlisting}

\begin{definition}[The axiom schema for induction]
This is the foundation of Peano arithmetic, its allows us to conclude a result holds for all natural numbers if an initial condition and a progression condition hold true. It is also considered the most suspect rule, as it is not finitary.\cite{chow}
\

Ind \ : \ $\forall \ A[x_i], \ \ A[x_i/0] \to (\forall m, A[x_i/m] \to A[x_i/m + 1]) \to \forall m, A[x_i/m]$
\end{definition}
\begin{lstlisting}[language=Coq,label=list:PeanoCoq,abovecaptionskip=-\medskipamount]
Inductive Peano_Theorem : formula -> Type :=
...
| induct : forall (A : formula) (n : nat),
      Peano_Theorem (substitution A n zero ->
          ((univ n (A -> (substitution A n (succ (var n))))) ->
              (univ n A))).
...
\end{lstlisting}

\begin{definition}[Deductive rules of Peano Arithmetic]
The deductive rules are what transform this from a set of axioms into a Hilbert Calculus. Modus Ponens (or MP) allows for implication elimination, deriving the conclusion of an implication from the implication and its premise. Universal generalisation introduces the binding of a free variable, following the logic that if a statement was true when the variable had no specific value then it must be true for all values.
\begin{description}
\item{MP :} \[\AxiomC{$A$}\AxiomC{$A \rightarrow B$}\BinaryInfC{$B$}\DisplayProof\]

\item{UG :} \[\AxiomC{$A[x_i]$}\UnaryInfC{$\forall m, A[x_i/m]$}\DisplayProof\]
\end{description}
\end{definition}

\begin{lstlisting}[language=Coq,label=list:PeanoCoq,abovecaptionskip=-\medskipamount]
Inductive Peano_Theorem : formula -> Type :=
...
| MP : forall (A B : formula),
                Peano_Theorem (A -> B) -> Peano_Theorem A -> Peano_Theorem B
| UG : forall (A : formula) (n : nat), Peano_Theorem A -> Peano_Theorem (univ n A) 
\end{lstlisting}

\

\subsection{Finitely Representable Ordinals}

\begin{definition}[Standard Representation of the Ordinal Numbers]
\
    \begin{itemize}
    \item 0 is an ordinal.
    \item If $\alpha$ is an ordinal then so is $S(\alpha)$.
    \item If $\beta_i$ are a countable set of ordinals such that $\beta_n < \beta_{n+1}$ then $\alpha = \lim\limits_{n\to \infty} \beta_n$ is an ordinal.
    \end{itemize}
  \end{definition}

The standard definition of the ordinals starts with the normal representation of the natural numbers, with 0 as a primitive element and the successor function which generates additional elements. The ordinal numbers definition adds an additional constructor, for constructing `limit ordinals'. This limit constructor takes a (countably) infinite set of ordinals and produces a new ordinal with the property that it is greater than all ordinals in the set used to construct it. The first of these limit ordinals, achieved by applying the limit constructor to all elements of $\mathbb{N}$ is called $\omega$. We can then apply the successor function to $\omega$ to produce $\omega + 1$ or indeed $\omega + n$ for any natural number $n$. With infinite repeated applications of these successor and limit constructors, we can generate more elements such as $2 \times \omega$, $\omega^2$ and even $\omega^\omega$. Continuing this process further will result in increasingly long exponential chains ($\omega^{\omega^{\omega^{\tiny{\cdot^{\cdot}}}}})$ of $\omega$. The limit ordinal which completes that process is known as $\epsilon_0$, we can continue the process beyond this point to make further ordinals, but they are not useful for this proof. As such for the remainder of this paper, when we refer to ordinals we mean ordinals less than $\epsilon_0$.

\

Such ordinals have the important property of being finitely representable. An important consequence of this is that it provides an alternate definition, since the standard definition of the ordinals involves defining $\omega$ in a manner comparably to $\mathbb{N}$, the set of all natural numbers. As such, using the standard definition, the introduction of the axiom that bounded induction up to $\epsilon_0$ is valid would trivially imply that unbounded induction over $\mathbb{N}$ was valid. There are two ways that we can think of the ordinals less than $\epsilon_0$ as finite, they can be represented by the structure of nested lists, or by finite trees.\footnote{It is important to note that $\epsilon_0$ itself is not finitely representable.}

\

\begin{definition}[Ordinals as Nested Lists]
\label{definition:ordinallists}
\
\begin{itemize}
  \item The empty list, is the ordinal 0.
  \item The list $\alpha$, with the $n$ entries $\alpha_1, \cdots, \alpha_n$, is the ordinal $\omega^{\alpha_1} + \cdots + \omega^{\alpha_n}$
\end{itemize}
\end{definition}

\noindent Therefore the empty list [] encodes the ordinal 0, while the list [[]] would encode $\omega^0 = 1$ and the list [[[]],[[]]] would be $\omega^{\omega^0} + \omega^{\omega^0} = \omega^1 + \omega^1 = \omega \times 2$.

\

\begin{definition}[Ordinals as Finite Trees]
\label{definition:ordinaltrees}
\
\begin{itemize}
  \item The trivial tree, is the ordinal 0.
  \item \begin{tabular}{p{6cm}c}
  \vspace*{-15mm}The tree $\alpha$, having a root node with $n$ branches $\alpha_1, \cdots, \alpha_n$, is the ordinal $\omega^{\alpha_1} + \cdots + \omega^{\alpha_n}$.&
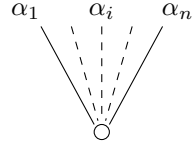
{\begin{tikzpicture}
  \draw (0,0) node {$\alpha_1$};
  \draw (1,0) node {$\alpha_i$};
  \draw (2,0) node {$\alpha_n$};
  \draw (1,-1.6) circle (0.1cm);
  \draw (0.2,-0.2) -- (0.9,-1.5);
  \draw (1.8,-0.2) -- (1.1,-1.5);
  \draw[dashed] (1.4,-0.2) -- (1.05,-1.45);
  \draw[dashed] (0.6,-0.2) -- (0.95,-1.45);
  \draw[dashed] (1,-0.2) -- (1,-1.4);
  \end{tikzpicture}}
  \end{tabular}
\end{itemize}
\end{definition}

These definitions are essentially equivalent, however the tree definition is usually easier to visualise, whereas the list version has more compact notation. For example, the notion of normal form for ordinals (which is used for defining their arithmatic properties) is the same as requiring that from any node of the tree the longest path from that node away from the root is the leftmost path. Other paths can be equal in length, but never longer. An observation that can be made at this point is that for any non-zero ordinal in normal form, the leftmost branch must be a power of $\omega$. Each other branch must be equal to this branch or strictly smaller. As such we can group all of those like branches, and instead represent our ordinal as a tuple of two ordinals and a natural number, where the first ordinal is the exponent of $\omega$ in the representation of the leftmost branch, the natural number is the number of additional copies of that branch, and the final ordinal represents the rest of the tree. As such the tuple $(\alpha,n,\beta)$ represents the ordinal $\omega^{\alpha} \times (n + 1) + \beta$.

\begin{definition}[Ordinals less than $\epsilon_0$ within Coq]
\label{definition:ordinalCoq}
\end{definition}
\begin{lstlisting}[language=Coq,label=list:ordinalCoq,abovecaptionskip=-\medskipamount]
Inductive ord : Set :=
| Zero : ord
| cons : ord -> nat -> ord -> ord.
\end{lstlisting}

\noindent The constructor cons implements the tuple mentioned above, i.e. cons alpha n beta = $\omega^\alpha \times (n+1) + \beta$.

\

Definition~\ref{definition:ordinalCoq} has the advantage of being a very simple inductive type, and does not need to invoke any set theory; however, it does come with some drawbacks. The main issue is the twin problem of constructing the well-ordering and defining the normal form for ordinals. Unlike the standard definition, \ref{definition:ordinalCoq} does not automatically define an order over the set of ordinals and furthermore can produce ordinals which are not in normal form. These problems then reinforce each other, because without an order it is impossible to define the normal form, yet it is only easy to provide an ordering for ordinals in normal form. The solution we came up with was borrowed from Pierre Cast\'eran and Evelyne Contejean's work on ordinals in Coq~\cite{casteran}. Instead of trying to define an ordering which behaves appropriately for
non-normal ordinals, we instead define an ordering which has the expected behaviour over ordinals in normal form. This is sufficient to allow us to define normal form, and we can now safely ignore its behaviour on the rest of the set, as from this point on we will restrict our attention to only consider ordinals in normal form.

\begin{definition}[Ordering over the Ordinals]
\label{definition:ordinalordering}
The ordering of the ordinals is the lexicographic ordering of the tuple representation.
\begin{itemize}
  \item The ordinal 0 is less than all ordinals of the form $(\alpha,n,\beta)$.
  \item If $\alpha_1$ is less than $\alpha_2$ then $(\alpha_1,n_1,\beta_1)$ is less than $(\alpha_2,n_2,\beta_2)$ for all $n_1,n_2,\beta_1,\beta_2$.
  \item If $\alpha_1 = \alpha_2$ and $n_1$ is less than $n_2$ then $(\alpha,n_1,\beta_1)$ is less than $(\alpha,n_2,\beta_2)$ for all $\beta_1,\beta_2$.
  \item If $\alpha_1 = \alpha_2,n_1 = n_2$ and $\beta_1$ is less than $\beta_2$ then $(\alpha,n,\beta_1)$ is less than $(\alpha,n,\beta_2)$. 
\end{itemize}
\end{definition}
\begin{lstlisting}[language=Coq,label=list:orderingCoq,abovecaptionskip=-\medskipamount]
Inductive ord_lt : ord -> ord -> Prop :=
| zero_lt : forall alpha n beta, Zero < cons alpha n beta
| head_lt : forall alpha1 alpha2 n1 n2 beta1 beta2 , alpha1 < alpha2 ->
                    cons alpha1 n1 beta1 < cons alpha2 n2 beta2
| coeff_lt : forall alpha n1 n2 beta1 beta2 , (n1 < n2 )%nat ->
                    cons alpha n1 beta1 < cons alpha n2 beta2
| tail_lt : forall alpha n beta1 beta2 , beta1 < beta2 ->
                    cons alpha n beta1 < cons alpha n beta2
where "alpha1 < alpha2 " := (ord_lt alpha1 alpha2).
\end{lstlisting}

\begin{definition}[Normal Form Ordinals]
\label{definition:ordinalnormal}
An ordinal is normal if its component ordinals are normal and it is correctly ordered.
\begin{itemize}
  \item The ordinal 0 is normal.
  \item The ordinal $(\alpha,n,0)$ is normal if $\alpha$ is normal.
  \item The ordinal $(\alpha,n,(\beta,m,\gamma))$ is normal if both $\alpha$ and $(\beta,m,\gamma)$ are normal and $\beta$ is less than $\alpha$.\footnote{While this definition may look slightly unintuitive, it is necessary to correctly evaluate the order of $\alpha$ and $\beta$ within $(\alpha,n,\beta) = \omega^\alpha \times (n+1) + \beta$. By also splitting the second term, we can compare the exponents of $\omega$ directly.  Even though it does look slightly strange, it is very simple to work with.}
\end{itemize}
\end{definition}
\begin{lstlisting}[language=Coq,label=list:normalCoq,abovecaptionskip=-\medskipamount]
Inductive nf : ord -> Prop :=
| zero_nf : nf Zero
| single_nf : forall alpha n,
              nf alpha -> nf (cons alpha n Zero)
| cons_nf : forall alpha n beta m gamma,
              beta < alpha -> nf alpha -> nf (cons beta m gamma) ->
                        nf (cons alpha n (cons beta m gamma)).
\end{lstlisting}

With the ordinal representation and their ordering complete, all that remains is to implement the necessary arithmetic operations and prove that they are closed under normal forms. Necessary functions include: successor, predecessor, addition, multiplication, exponentiation and maximum. Unlike with the ordering, there are not multiple possible approaches here and so the details of the implementation and closure of the operations will be omitted for brevity. There are also a great deal of monotonicity proofs necessary to complete the proof, however these are all procedural and not instructive.

The only remaining proof of interest relating to the ordinals is the proof that strong induction is valid. Technically, this could have been added as an axiom within Coq rather than proved, as it is an axiom of PRA + $\epsilon_0$, which is the system which we are trying to work within. However, since Cast\'eran and Contejean had already proven that strong induction held, we reimplimented their method to prove the necessary result.

\begin{theorem}[Strong Induction over the Ordinals]
Given an ordinal $\alpha$ in normal form. If a proposition $P$ holding for every normal form ordinal less than $\alpha$ implies that the proposition holds for $\alpha$. Then we can conclude that $P$ holds for all ordinals in normal form.
\end{theorem}

\

\begin{lstlisting}[language=Coq,label=list:inductionCoq,abovecaptionskip=-\medskipamount]
Theorem transfinite_induction :
    forall (P: ord -> Type),
        (forall x: ord, nf x ->
            (forall y: ord, nf y -> ord_lt y x -> P y) -> P x) ->
        forall a, nf a -> P a.
\end{lstlisting}

\subsection{The PA$_\omega$ Sequent Calculus}

PA$_\omega$ is the sequent calculus that Gentzen invented for this consistency proof. Its language contained the initial symbol $0$, a countable set of free variables $x_0, x_1 \cdots, x_n, \cdots$ the unary symbol $S$, the binary symbols $+,\times,=$ and the logical symbols $\neg,\vee$ and $\forall$. equipped with the following grammar.

\begin{definition}[The Language of PA$_\omega$]
\label{definition:BNF}
\

$term \ := \ 0 \ | \ x_i \ | \ S (term) \ | \ term + term \ | \ term \times term$

$atom \ := \ term = term$

$\phi \ := \ atom \ | \ \neg \phi \ | \ \phi \vee \phi \ | \ \forall x_n, \phi$
\end{definition}
\begin{lstlisting}[language=Coq,label=list:languageCoq,abovecaptionskip=-\medskipamount]
Inductive term : Type :=
| zero : term
| succ : term -> term
| plus : term -> term -> term
| times : term -> term -> term
| var : nat -> term.

Inductive atomic_formula : Type :=
| equ : term -> term -> atomic_formula.

Inductive formula : Type :=
| atom : atomic_formula -> formula
| neg : formula -> formula
| lor : formula -> formula -> formula
| univ : nat -> formula -> formula.
\end{lstlisting}

\begin{example}[Formulae in PA$_\omega$ and their Coq Representations]
\label{example:formulae}
\
\begin{itemize}
  \item $0 + x_0 = x_1 \times S ( 0 )$ \ \ {\normalfont \lstinline[label=list:example1Coq,abovecaptionskip=-\medskipamount]{atom (equ (plus zero (var 0)) (times (var 1) (succ zero)))}}
  \item $ 0 = S ( 0 ) \vee 0 = 0$ \ \ {\normalfont \lstinline[label=list:example1Coq,abovecaptionskip=-\medskipamount]{lor (atom (equ zero (succ zero))) (atom (equ zero zero))}}
  \item $ \neg ( \forall x_5, x_5 = 0 )$ \ \ {\normalfont \lstinline[label=list:example1Coq,abovecaptionskip=-\medskipamount]{neg (univ 5 (atom (equ (var 5) zero)))}}
\end{itemize}
\end{example}

The initial sequents for the PA$_\omega$ calculus are all quantifier-free atomic formulae which are true, and the negations of quantifier-free atomic formulae that are false. Any atomic formula which does not contain free variables has a deterministic truth value, which can be checked by simply evaluating each side of the equality. If the equality holds then the formulae representing it will be an initial sequent of PA$_\omega$, and if it does not then its negation will be an initial sequent instead. This determination of the truth value can be performed within Coq with an evaluation function, which returns a 0 if the term contains any free variables and 1 greater than its true value otherwise.\footnote{This encoding is useful as it avoids the need for defining a maybe type or product type, and the only downside is that the evaluation constant is offset from the true value. However, this is rarely relevant and is easily accounted for when necessary.}

\begin{example}[The Evaluation Function] A quantifier free atomic formula evaluates to one higher than its arithmetic value, atomic formula containing quantifiers evaluate to 0.
\label{exampe:evaluations}
\begin{itemize}
  \item The evaluation of $S ( 0 )$ is 2.
  \item The evaluation of $( 0 \times ( S ( S ( 0 ) ) )$ is 1.
  \item The evaluation of $S (x_7)$ is 0.
\end{itemize}
\end{example}

It is worth noting that given this construction, PA$_\omega$ is not finitely presented. However, it does only have a countably infinite number of initial sequents, which will be relevant for one of the deductive rules.  It is useful to break up the deductive rules of PA$_\omega$ into 3 categories, the short rules, the tall rules, and Cut. The short rules are Exchange and Contraction, and they do not change the logical content of a sequent, only its form. The tall rules are Weakening, Negation, Quantification, DeMorgan, and $\omega$. These rules each introduce or modify one of the logical symbols in the language $\vee,\neg\neg,\neg\forall,\neg\vee$ and $\forall$ respectively. The final category is cut, which contains the cut rule, which is
used to eliminate contradictions.

\begin{figure}
\begin{definition}[The Deductive Rules of PA$_\omega$]
\label{def:PA_rules}
\phantom{a}\\
\begin{tabular}
  {|cc|}
  \hline
  & \\
  \multicolumn{2}{|c|}{\large{\textbf{Initial}}}\\
  \hspace*{0.03\linewidth}
  \AxiomC {}
  \RightLabel{$Atom_t$}
  \UnaryInfC {$\vdash t_1 = t_2$}
  \DisplayProof
  \hspace*{0.05\linewidth}
  &
  \AxiomC {}
  \RightLabel{$Atom_f$}
  \UnaryInfC {$\vdash \neg (t_1 = t_2)$}
  \DisplayProof\\
  If eval($t_1$) = eval($t_2$) $\ne 0$ & If $0 \ne$ eval($t_1$) $\ne$ eval($t_2$) $\ne 0$ \\
  & \\
  \hline
  & \\
  \multicolumn{2}{|c|}{\large{\textbf{Short}}}\\
  \hspace*{0.03\linewidth}
  \AxiomC {$((C \vee A) \vee B) \vee D$}
  \RightLabel{Exchange}
  \UnaryInfC {$((C \vee B) \vee A) \vee D$}
  \DisplayProof
  \hspace*{0.05\linewidth}
  &
  \AxiomC {$(A \vee A) \vee D$}
  \RightLabel{Contraction}
  \UnaryInfC {$A \vee D$}
  \DisplayProof\\
  & \\
  \hline
  & \\
  \multicolumn{2}{|c|}{\large{\textbf{Tall}}}\\
  \AxiomC {$D$}
  \RightLabel{Weakening}
  \UnaryInfC {$A \vee D$}
  \DisplayProof
  &
  \AxiomC {$A \vee D$}
  \RightLabel{Negation}
  \UnaryInfC {$\neg \neg A \vee D$}
  \DisplayProof\\
  & \\
  \AxiomC {$\neg(A[x_i/m]) \vee D$}
  \RightLabel{Quantification}
  \UnaryInfC {$\neg(\forall t, A[x_i/t]) \vee D$}
  \DisplayProof
  &
  \AxiomC {$\neg (A) \vee D$}
  \AxiomC {$\neg (B) \vee D$}
  \RightLabel{DeMorgan}
  \BinaryInfC {$\neg(A \vee B) \vee D$}
  \DisplayProof\\
  & \\
  \multicolumn{2}{|c|}{
  \hspace*{0.01\linewidth}
  \AxiomC {$A[x_i/0] \vee D$ \hspace*{0.4cm} $A[x_i/1] \vee D$ \hspace*{0.3cm} $\cdots$}
  \AxiomC {$A[x_i/m] \vee D$ \hspace*{0.3cm} $\cdots$ \hspace*{0.2cm}}
  \RightLabel{$\omega$}
  \BinaryInfC {$(\forall t, A[x_i/t]) \vee D$}
  \DisplayProof
  \hspace*{0.01\linewidth}}\\
  & \\
  \hline
  & \\
  \multicolumn{2}{|c|}{\large{\textbf{Cut}}}\\
  & \\
  \multicolumn{2}{|c|}{  
  \AxiomC {$C \vee A$}
  \AxiomC {$\neg A \vee D$}
  \RightLabel{Cut}
  \BinaryInfC {$C \vee D$}
  \DisplayProof}\\
  & \\
  \hline
\end{tabular}
\end{definition}
\

\noindent In all rules except Weakening the existence of the side-formula $D$ is optional, and in all rules the existence of the side-formula $C$ is optional, however for the cut rule at least one of $C$ and $D$ must exist.

\

\noindent Where the notation $A[x_i/m]$ means replacing all unbound instances of $x_i$ within A with the term $m$.
\end{figure}

\

The reason for the classification of the rules, is the decorations that will be applied to the sequents which we need for the proof. The decorations include a natural number called the degree and an ordinal which is called the height. A short rule does not effect either the degree or the height of a derivation. A tall rule does not increase the degree of a derivation but does increase its height. The cut rule always increases the height of a derivation, and may increase the degree of the derivation, depending on the size of the cut formula relative to the degree. As usual, we define a derivation as follows:

\begin{definition}[Derivation]
\begin{itemize}
\
  \item Every initial sequent is a derivation of height $0$ and degree 0.
  \item If the endsequent of a derivation $D$ with height $\alpha$ and degree $d$ matches the premise of a short rule $R$, then extending the derivation by applying $R$ to $D$ gives a new derivation with height $\alpha$ and degree $d$.
  \item If the endsequents of a set of derivations $D_i$ with heights $\alpha_i$ and degree $d_i$ match the set of premises of a strong rule $R$ then extending the derivation by applying $R$ to the set $D_i$ gives a new derivation with height $\sup\{\alpha_i\}+1$ and degree $\max\{d_i\}$. For this to make sense, this enforces a side condition for the $\omega$-rule, that there must be some maximum degree among all its infinite premises and that the set of heights must contain a supremum which is itself a finitely representable ordinal, (i.e. the supremum can not be $\epsilon_0$.
  \item If the endsequents of two derivations $D_1$ and $D_2$ with heights $\alpha_1$ and $\alpha_2$ and degrees $d_1$ and $d_2$ match the premises of a cut rule over the formula $\phi$ then extending the derivation by applying the cut rule to $\{D_1,D_2\}$ gives a new derivation with degree $\max\{d_1,d_2,|\phi|+1\}$ (where $|\phi|$ is defined as the number of logical connectives in $\phi$) and with height $\max\{\alpha_1,\alpha_2\}+1$ if $C$ exists or height $\max\{\alpha_1,\alpha_2\}+2$ if it does not\footnote{This is a special case to take into account a workaround for a technique that relies on side formulae being present and will be discussed later.}.
\end{itemize}
\end{definition}

\

When translating this construction into Coq, it becomes necessary to explicitly deal with side formulae. While this is often done with the use of list operations, that is inconvenient for the definition of formula we are using. Instead we duplicate each constructor, to have a different form for each combination of side formulae. This can be seen with the 3 versions of the cut rule at the end of the sample code.

\

\begin{definition}[Coq Construction of PA$_\omega$ with Decorations]
The theorems of PA$_\omega$ can be built using the following inductive construction.
\end{definition}
\begin{lstlisting}[language=Coq,label=list:theoremCoq,abovecaptionskip=-\medskipamount]
Inductive PA_omega_theorem : formula -> nat -> ord -> Type :=
| axiom      : forall (A : formula), PA_omega_axiom A = true ->
                PA_omega_theorem A 0 Zero

| exchange1  : forall (A B : formula) (d : nat) (alpha : ord),
               PA_omega_theorem (lor A B) d alpha ->
                PA_omega_theorem (lor B A) d alpha

(*Other deductive rule implementations omitted for brevity*)

| w_rule2 : forall (A D : formula) (n : nat) (d : nat) (alpha : ord)
          (g : forall (t : term), closed_t t = true ->
            PA_omega_theorem (lor (substitution A n t) D) d alpha),
          PA_omega_theorem (lor (univ n A) D) d (ord_succ alpha)

| cut1: forall (C A : formula) (d1 d2 : nat) (alpha1 alpha2 : ord),
        PA_omega_theorem (lor C A) d1 alpha1 ->
        PA_omega_theorem (neg A) d2 alpha2 ->
        PA_omega_theorem C (max (max d1 d2) (num_conn (neg A)))
              (ord_succ (ord_max alpha1 alpha2))
| cut2: forall (A D : formula) (d1 d2 : nat) (alpha1 alpha2 : ord),
        PA_omega_theorem A d1 alpha1 ->
        PA_omega_theorem (lor (neg A) D) d2 alpha2 ->
        PA_omega_theorem D (max (max d1 d2) (num_conn (neg A)))
            (ord_succ (ord_succ (ord_max alpha1 alpha2)))
| cut3: forall (C A D : formula) (d1 d2 : nat) (alpha1 alpha2 : ord),
        PA_omega_theorem (lor C A) d1 alpha1 ->
        PA_omega_theorem (lor (neg A) D) d2 alpha2 ->
        PA_omega_theorem (lor C D) (max(max d1 d2) (num_conn(neg A)))
            (ord_succ (ord_max alpha1 alpha2)).
\end{lstlisting}

\

With this definition it is then also possible to prove that PA$_\omega$ admits the normal associativity rules, and that the ordinals associated with a theorem are in normal form. Both of these facts will become critical in the cut elimination argument that follows.

\

\begin{lemma}[PA$_\omega$ is Associative]

A formula $((C \vee A) \vee B)$ is a theorem of PA$_\omega$ if and only if $(C \vee (A \vee B))$ is a theorem of PA$_\omega$.
\end{lemma}
\begin{lstlisting}[language=Coq,label=list:assocCoq,abovecaptionskip=-\medskipamount]
Lemma associativity1  : forall (C A B : formula) (d : nat) (alpha : ord),
                        PA_omega_theorem (lor (lor C A) B) d alpha ->
                        PA_omega_theorem (lor C (lor A B)) d alpha.

Lemma associativity2  : forall (C A B : formula) (d : nat) (alpha : ord),
                        PA_omega_theorem (lor C (lor A B)) d alpha ->
                        PA_omega_theorem (lor (lor C A) B) d alpha.
\end{lstlisting}

\

\begin{lemma}[PA$_\omega$ Derivations Have Normal Height]

If $A$ is a theorem of PA$_\omega$ with a derivation of height $\alpha$, then $\alpha$ is an ordinal in normal form.
\end{lemma}
\begin{lstlisting}[language=Coq,label=list:theoremnfCoq,abovecaptionskip=-\medskipamount]
Lemma theorem_nf  : forall (A : formula) (d : nat) (alpha : ord),
                        PA_omega_theorem A d alpha ->
                        nf alpha.
\end{lstlisting}
s
\subsection{Derivations in PA$_\omega$}

While the \lstinline{PA_omega_theorem} construct represents that a formula $A$ can be derived in PA$_\omega$ with degree $d$ and height $\alpha$ using only the axioms and deductive rules, objects of this type do not contain the entire derivation, only the final step. However, many of the proofs will involve searching through more than just the final step of a derivation, especially cut elimination. To this end a second type is built which will contain the additional information of every deductive rule in the proof, which we shall call a \lstinline{ptree} (short for proof tree). The \lstinline{ptree} type is much less convenient to work with since it has many more parts, which justifies the existence of \lstinline{PA_omega_theorem}, which shall be used for any results which do not need this extra structure. In effect a \lstinline{ptree} will be a tree structure which takes as parameters all of the information describing the current deductive step, and proof trees for each of its premises.

\begin{definition}[Proof Tree Construction]
\label{def:prooftree}
The structure of a proof tree is defined inductively as follows. For each deductive rule in PA$_\omega$ we will assign a constructor for the proof tree. This constructor will take as parameters, one formula for each principle formulae of the deductive rule, one formula for each side formula of the rule, a natural number for each premise of the rule, an ordinal for each premise of the rule, and one proof tree for each premise of the rule. Additionally the constructor for quantification and the $\omega$-rule will also take a further natural number as a parameter to reference the free variable in question. 
\end{definition}
\begin{lstlisting}[language=Coq,label=list:ptreeCoq,abovecaptionskip=-\medskipamount]
Inductive ptree : Type :=
| deg_up : nat -> ptree -> ptree
| ord_up : ord -> ptree -> ptree
| node : formula -> ptree
| exchange_ab : formula -> formula -> nat -> ord -> ptree -> ptree
| exchange_cab : formula -> formula -> formula ->
                      nat -> ord -> ptree -> ptree
| exchange_abd : formula -> formula -> formula ->
                      nat -> ord -> ptree -> ptree
| exchange_cabd : formula -> formula -> formula -> formula ->
                      nat -> ord -> ptree -> ptree
| contraction_a : formula -> nat -> ord -> ptree -> ptree
| contraction_ad : formula -> formula -> nat -> ord -> ptree -> ptree
| weakening_ad : formula -> formula -> nat -> ord -> ptree -> ptree
| demorgan_ab : formula -> formula ->  nat -> nat -> ord -> ord ->
                      ptree -> ptree -> ptree
| demorgan_abd : formula -> formula -> formula -> nat -> nat ->
                      ord -> ord -> ptree -> ptree -> ptree
| negation_a : formula -> nat -> ord -> ptree -> ptree
| negation_ad : formula -> formula -> nat -> ord -> ptree -> ptree
| quantification_a : formula -> nat -> term -> nat ->
                      ord -> ptree -> ptree
| quantification_ad : formula -> formula -> nat -> term -> nat ->
                      ord -> ptree -> ptree
| w_rule_a : formula -> nat -> nat -> ord ->
                      (c_term -> ptree) -> ptree
| w_rule_ad : formula -> formula -> nat -> nat -> ord ->
                      (c_term -> ptree) -> ptree
| cut_ca : formula -> formula ->  nat -> nat -> ord -> ord ->
                      ptree -> ptree -> ptree
| cut_ad : formula -> formula ->  nat -> nat -> ord -> ord ->
                      ptree -> ptree -> ptree
| cut_cad : formula -> formula -> formula -> nat -> nat ->
                      ord -> ord -> ptree -> ptree -> ptree.  
\end{lstlisting}

\begin{example}[A Simple Proof Tree] We provide here a deduction of $(0 + 0 = 0) \vee (x_1 = 0)$ and its representation within our Coq formulation.

\

\AxiomC {$(0 + 0 = 0)$}
\RightLabel{Weakening}
\UnaryInfC {$(x_1 = 0) \vee (0 + 0 = 0)$}
\RightLabel{Exchange}
\UnaryInfC {$(0 + 0 = 0) \vee (x_1 = 0)$}
\DisplayProof
\end{example}

\begin{lstlisting}[language=Coq,label=list:exampleptree,abovecaptionskip=-\medskipamount]
P : ptree :=
    exchange_ab
        (atom (equ (var 1) zero))
        (atom (equ (plus zero zero) zero))
        0
        (cons Zero 0 Zero)
        (weakening_ad
            (atom (equ (plus zero zero) zero))
            (atom (equ (var 1) zero))
            0
            Zero
            (node (atom (equ (plus zero zero) zero))))
\end{lstlisting}

After having defined these \lstinline{ptree}'s it is necessary to create functions to extract the relevant information from them. To this end a series of functions \lstinline{ptree_formula}, \lstinline{ptree_deg} and \lstinline{ptree_ord} are defined. The implementation of these functions is done via a standard pattern matching method. With this functions, it is then possible to define a predicate on \lstinline{ptree}'s that ensures that they are well-formed. This acts as a catch-all function which enforces any necessary side conditions, checks that the height and degree change as expected, as well as ensuring that the principle formula is present in the correct location and that the side formulae are unchanged. Otherwise, the constructors of \lstinline{ptree} would have no guard conditions, so most things of type \lstinline{ptree} do not represent a sensible series of deductions, whereas well-formed \lstinline{ptree}'s will have corresponding \lstinline{PA_Omega_Theorem} constructions.\footnote{This represents another advantage of the \lstinline{ptree} type, as it allows us to define the transformation as we wish, and then later check that it satisfies the necessary properties. The \lstinline{PA_Omega_Theorem} type does not allow this flexibility.} The well-formed predicate performs a series of checks to make sure that all of the parameters are of the correct form for the deductive rule being applied.

\begin{definition}[Well-formed Proof Trees]
If a proof tree is well-formed or not can be determined inductively. A node is well-formed if the formula is an initial sequent. A constructor associated with a deductive rule is well-formed if:
\begin{itemize}
  \item All proof trees given as arguments are well-formed. These will be called the premises.
  \item Each premise has a degree with the same value as the associated natural number parameter
  \item Each premise has a height with the same value as the associated ordinal parameter
  \item Each premise has a formula with the correct structure for the application of the rule in question, and the subformulae of this formula are correctly described by the associated formula parameters.
  \item All formula parameters are closed.\footnote{For all except \lstinline{weakening_ad} and \lstinline{node} this is implicit.}
\end{itemize}
\end{definition}

\begin{definition}[Provable Formulae]
A formula $A$ is provable at $(d,\alpha)$ for $d \in \mathbb{N}$ and $\alpha < \epsilon_0$ if there exists some well-formed \lstinline{ptree} $P$ with degree $d$, height $\alpha$ and formula $A$.
\end{definition}

\begin{lstlisting}[language=Coq,label=list:formula_provable,abovecaptionskip=-\medskipamount]
Definition provable (A:formula) (d:nat) (alpha:ord) : Type :=
    {P : ptree & (ptree_formula P = A) * (well_formed P) *
          (d >= ptree_deg P) * (ptree_ord P = alpha)}.
\end{lstlisting}

With these predicates defined, it is possible to prove that the subset of well-formed proof trees is in correspondence with set \lstinline{PA_Omega_Theorem}.

\begin{theorem}[All Theorems are Provable]
If the formula $A$ is a theorem of PA$_\omega$ with degree $d$ and height $\alpha$, then there is a well-formed proof tree of degree $d$ and height $\alpha$ which has $A$ as as its conclusion.   
\end{theorem}

\begin{lstlisting}[language=Coq,label=list:provable_theorem,abovecaptionskip=-\medskipamount]
Theorem provable_theorem: forall (A : formula)(d : nat)(alpha : ord),
  PA_omega_theorem A d alpha -> provable A d alpha.
\end{lstlisting}

\begin{theorem}[All Proof Trees Prove Theorems]
If $P$ is a well-formed proof tree with conclusion $A$, degree $d$ and height $\alpha$, then $A$ is a theorem of PA$_\omega$ with degree $d$ and height $\alpha$.
\end{theorem}
\begin{lstlisting}[language=Coq,label=list:theorem_provable,abovecaptionskip=-\medskipamount]
Theorem theorem_provable: forall (A : formula)(d : nat)(alpha : ord),
  provable A d alpha -> PA_omega_theorem A d alpha.
\end{lstlisting}

The final necessary component to manipulate proof trees is defining a substitution schema over proof trees. The substitution schemas represent a method to systematically change instances of a formula $A$ within a \lstinline{ptree} to instead be some other formula $B$. In general this does not represent a logical operation, however it does provide a way to define transformations of \lstinline{ptree}s which do not change the structure of the sequent. Therefore, we will be able to show that if the replacement was logically sound when it first occurred in the derivation, the transformed \lstinline{ptree} will also be well-formed. The primary difficulty in defining a substitution schema that can preserve the form in such a way stems from the potential for multiplicity of formulae. Due to the presence of contraction and weakening, the number of instances of the formula to be replaced ($A$), can fluctuate as the derivation progresses, and the presence of exchange means that the position of these instances relative to each other may not be fixed either. If it was known that all instances of the formula $A$ needed to be replaced, this would not be a problem, but this is too strong a claim, since the tall rules may transform one copy of $A$ to some third formula $C$, meaning that any $A$ in the derivation history of $C$ needs to be left unchanged.

\begin{example}
\label{example:Meta Substitution}
A worked example of naive substitution failing to be logically sound.

\

$\begin{array}{ccc}
\ \ \ \ \mathrm{Original \ Derivation \ \ \ \ } & \mathrm{Desired \ Outcome \ \ \ \ \ \ } & \mathrm{Naive \ Substitution}\\
\AxiomC {$A$}
\LeftLabel{\hspace*{-1cm}\tiny{Weakening}}
\UnaryInfC{$A \vee A$}
\LeftLabel{\hspace*{-1cm}\tiny{Tall Rule}}
\UnaryInfC {$C \vee A$}
\DisplayProof &
\AxiomC {$B$}
\LeftLabel{\hspace*{-2cm}\tiny{Weakening}}
\UnaryInfC{$A \vee B$}
\LeftLabel{\hspace*{-2cm}\tiny{Tall Rule}}
\UnaryInfC {$C \vee B$}
\DisplayProof &
\AxiomC {$B$}
\LeftLabel{\hspace*{-2cm}\tiny{Weakening}}
\UnaryInfC{$B \vee B$}
\LeftLabel{\hspace*{-2cm}\tiny{Tall Rule}}
\RightLabel{\tiny{Does not follow since $A \ne B$ \hspace*{-4cm}}}
\UnaryInfC {$C \vee B$}
\DisplayProof\\
\end{array}$
\end{example}

\

To this end we define a substitution indicator, which mimics the disjunctive structure of formulae and indicates which branches are to be considered for substitution and which should be ignored. 

\begin{definition}[Substitution Indicators]
A substitution indicator is a boolean tree. A given substitution indicator $S$ is said to fit a formula $A$ if the tree structure of $S$ matches the disjunctive structure of $A$.
\end{definition}

\begin{lstlisting}[language=Coq,label=list:substindCoq,abovecaptionskip=-\medskipamount]
Inductive subst_ind : Type :=
| ind_0 : subst_ind
| ind_1 : subst_ind
| lor_ind : subst_ind -> subst_ind -> subst_ind.

Fixpoint subst_ind_fit (A : formula) (S : subst_ind) : bool :=
match A, S with
| lor B C , lor_ind S_B S_C => subst_ind_fit B S_B && subst_ind_fit C S_C
| _ , lor_ind _ _  => false
| lor _ _ , _ => false
| _ , _ => true
end.
\end{lstlisting}

\

\begin{example}[Indicators Fitting]
The indicators (t) and (f) will fit any atomic formula, or any formulae of the form $\neg B$ or $\forall x_n, B(x_n)$, whereas the indicator $(S_1,(S_{21},S_{22}))$ will fit any formula of the form $B_1 \vee (B_{21} \vee B_{22})$, provided that $S_i$ fits $B_i$ for each $i$.
\end{example}

Because the substitution indicators mirror the structure of formulae, we can consider what effect applying the deductive rules upwards would have on the indicator. For any of the tall rules or cut, we will necessarily mask the principle formulae moving upwards, since necessarily it can only match the formula being replaced before or after the application of the rule, not both. Meanwhile the parts of the indicator relating to the side formulae will remain unchanged. For exchange, it will just permute the indicator the same way it would the formulae, and contraction will duplicate the indicator element associated with the formula being duplicated. With this in hand we now have a sensible framework for defining any of these substitution schemas.

\subsection{Law of the Excluded Middle with a Bound}
\label{sec:LEM}

Because the sequent calculus PA$_\omega$ includes deductive rules which account for double-negation and one of the DeMorgan identities the calculus describes a classical logic. As such, the calculus must admit the Law of the Excluded Middle, that is for any given closed formula $A$, it must be possible to construct a well-formed derivation which ends with the formula $\neg A \vee A$. As well as being a good check that the calculus has its expected properties, this construction is immensely valuable in proving the equivalence of Peano Arithmetic and PA$_\omega$. Most of the First Order Logic axioms of Peano Arithmetic can be derived using a carefully selected Law of the Excluded Middle formula and some applications of the demorgan deductive rule. The Law of the Excluded Middle is also essential for the derivation of the induction axiom, which introduces an additional constraint. Because the induction axiom involves the use of $\forall$, its derivation must by definition use the $\omega$-rule. However, the $\omega$-rule necessitates that all of the premises must have the same degree and a bound on the height. As such, it is insufficient to show that derivations of the formula $\neg A \vee A$ simply exist, but rather that they exist with a uniform degree and some calculable height which could be shown to be bounded over the premise set of the $\omega$-rule.

\

While the process for constructing the derivation is relatively straightforward, and is well documented, the matter of calculating the heights of these derivations did not seem to be explored. The process works by induction on the number of logical connectives in the formula $A$, and then breaking into cases based on what the outermost logical connective is. If there are no logical connectives, then since $A$ is closed, exactly one of the formulae $A$ and $\neg A$ are axioms of PA$_\omega$. As such weakening by the other formula and then potentially using exchange is enough to derive the desired formula in this case. If the final logical connective is negation, then a single application of exchange and negation to the inductive hypothesis is sufficient. If the final connective is disjunction, then the inductive hypothesis must be applied to each subformula, weakened by the other subformula and then combined using the demorgan rule after some exchanges. Finally, if the connective was $\forall$, the inductive hypothesis is instantiated for each premises, has quantification applied to it, then exchanged so that the $\omega$-rule applies, before being exchanged again to be in the correct form. Importantly, each step involves at most two uses of the tall rules in sequence (ignoring parallel applications, as multiple tall rules being performed in parallel does not increase height further than any one individual instance would). As such, no symbol can increase the height by more than 2, and all derivations were at least height 1, which gives us a final bound of $(2n + 1)$, where $n$ is the number of logical connectives in the formula $A$. Showing that the degree is uniform is trivial, since at no step is the cut rule used, so all of these derivations have degree 0. The most interesting element of this bound, is that it is finite, which has further implications for the proof of equivalence between Peano Arithmetic and PA$_\omega$.

\begin{center}
\begin{tabular}{|c|c|}
\hline
& \\
Case & Derivation of LEM\\
& \\
\hline
& \\
Initial $A$ &
\AxiomC{$A$}
\LeftLabel{\scriptsize$0, 1$}
\RightLabel{Weakening $\neg A$}
\UnaryInfC{$\neg A \vee A$}
\DisplayProof
\\
& \\
\hline
& \\
Initial $\neg A$ &
\AxiomC{$\neg A$}
\LeftLabel{\scriptsize$0, 1$}
\RightLabel{Weakening $A$}
\UnaryInfC{$A \vee \neg A$}
\LeftLabel{\scriptsize$0, 1$}
\RightLabel{Exchange 1}
\UnaryInfC{$\neg A \vee A$}
\DisplayProof
\\
& \\
\hline
& \\
Negation &
\AxiomC{$\vdots$}
\LeftLabel{\scriptsize$0, \alpha$}
\RightLabel{LEM ($A$)}
\UnaryInfC{$\neg A \vee A$}
\LeftLabel{\scriptsize$0, \alpha$}
\RightLabel{Exchange 1}
\UnaryInfC{$A \vee \neg A$}
\LeftLabel{\scriptsize$0, \alpha+ 1$}
\RightLabel{Negation 2}
\UnaryInfC{$\neg\neg A \vee \neg A$}
\DisplayProof
\\
& \\
\hline
& \\
Disjunction &
\scalebox{0.6}{
\AxiomC{$\vdots$}
\LeftLabel{\scriptsize$0, \alpha$}
\RightLabel{LEM ($A$)}
\UnaryInfC{$\neg A \vee A$}
\LeftLabel{\scriptsize$0, \alpha+1$}
\RightLabel{Weakening B}
\UnaryInfC{$B \vee (\neg A \vee A)$}
\LeftLabel{\scriptsize$0, \alpha+ 1$}
\RightLabel{Exchange 1}
\UnaryInfC{$(\neg A \vee A) \vee B$}
\LeftLabel{\scriptsize$0, \alpha+ 1$}
\RightLabel{Associativity 1}
\UnaryInfC{$\neg A \vee (A \vee B)$}

\AxiomC{$\vdots$}
\LeftLabel{\scriptsize$0, \beta$}
\RightLabel{LEM ($B$)}
\UnaryInfC{$\neg B \vee B$}
\LeftLabel{\scriptsize$0, \beta+1$}
\RightLabel{Weakening A}
\UnaryInfC{$A \vee (\neg B \vee B)$}
\LeftLabel{\scriptsize$0, \beta+ 1$}
\RightLabel{Exchange 1}
\UnaryInfC{$(\neg B \vee B) \vee A$}
\LeftLabel{\scriptsize$0, \beta+ 1$}
\RightLabel{Exchange 2}
\UnaryInfC{$(\neg B \vee A) \vee B$}
\LeftLabel{\scriptsize$0, \beta+ 1$}
\RightLabel{Associativity 1}
\UnaryInfC{$\neg B \vee (A \vee B)$}

\LeftLabel{\scriptsize$0, \mathrm{ord\_max}\{\alpha+1,\beta+ 1\}+1$}
\RightLabel{DeMorgan 2}
\BinaryInfC{$\neg (A \vee B) \vee (A \vee B)$}

\DisplayProof}
\\

& \\
\hline
& \\
Universal &

\scalebox{0.6}{
\LeftLabel{\scriptsize$0, \alpha$}
\AxiomC{$\vdots$}
\UnaryInfC{$\neg A[x_n/0] \vee A[x_n/0]$}
\LeftLabel{\scriptsize$0,\alpha+1$}
\UnaryInfC{$\neg (\forall t,A[x_n/t]) \vee A[x_n/0]$}
\LeftLabel{\scriptsize$0,\alpha+1$}
\UnaryInfC{$A[x_n/0] \vee \neg (\forall t,A[x_n/t])$}
\AxiomC{$\vdots$ \ \ \phantom{$A^{x^x}$} \ $\vdots$ \ \phantom{$A^{x^x}$} \ \ $\vdots$}
\UnaryInfC{$\cdots$ \phantom{$A^{x^x}$} $\cdots$ \phantom{$A^{x^x}$} $\cdots$}
\UnaryInfC{$\cdots$ \phantom{$A^{x^x}$} $\cdots$ \phantom{$A^{x^x}$} $\cdots$}
\UnaryInfC{$\cdots$ \phantom{$A^{x^x}$} $\cdots$ \phantom{$A^{x^x}$} $\cdots$}
\AxiomC{$\vdots$}
\RightLabel{LEM $(A[x_n/m])$}
\UnaryInfC{$\neg A[x_n/m] \vee A[x_n/m]$}
\RightLabel{Quantification 2}
\UnaryInfC{$\neg (\forall t,A[x_n/m]) \vee A[x_n/m]$}
\RightLabel{Exchange 1}
\UnaryInfC{$A[x_n/m] \vee \neg (\forall t,A[x_n/t])$}
\LeftLabel{\scriptsize$0,\alpha+2$}
\RightLabel{$\omega$-rule 2}
\TrinaryInfC{$(\forall t,A[x_n/t]) \vee \neg(\forall t,A[x_n/t])$}
\LeftLabel{\scriptsize$0,\alpha+2$}
\RightLabel{Exchange 1}
\UnaryInfC{$\neg (\forall t,A[x_n/t]) \vee (\forall t,A[x_n/t])$}
\DisplayProof}
\\
& \\
\hline
\end{tabular}
\end{center}

\

As you can see from the example above, while the invocation of other instances of LEM and the weakening are the only logical relevant steps, the full derivations contain a great many instances of exchange\footnote{Recalling that each associativity result is also just a specific sequence of exchanges}. This greatly expands the size of many derivations in this paper to an almost unworkable extent. So going forward, in order to strike a balance between completeness of the derivations and legibility of the proof trees, we will not show each step of the exchanges and associativity, and simply annotate the whole sequence as a single step. In the example above, the last 3 rules in the right branch of the disjunction case would be collapsed into a single line labelled $E_{1,2}A_1$, while the last 2 rules in the left branch would be $E_1A_1$.

\

As well as this normal formulation of the Law of the Excluded Middle, we can make a slightly more general statement which is that the Law applies not only reflexively to formula, but also two pairs of formula $A_1, A_2$ which are the same up to substitution of closed terms with equal evaluations. The proof of this statement is nearly identical to the the previous one, with the exception of the base case for atomic values, which instead uses the evaluation function to deduce the validity of the two potential initial sequents, instead of equality of formulae.

\begin{theorem}[Extended Law of the Exclued Middle]
\label{theorem:ELEM}
For all closed formulae $A_1$ and $A_2$ and closed terms $t_1$ and $t_2$, if $A_1 = A_2[t_2/t_1]$ and $(t_1 = t_2)$ is an initial sequent of PA$_\omega$, then $\neg A_1 \vee A_2$ is a theorem of PA$_\omega$.
\end{theorem}

\begin{lstlisting}[language=Coq,label=list:ELEM_Coq,abovecaptionskip=-\medskipamount]
Lemma LEM_term :
  forall (A : formula) (n : nat) (s t : term),
    correct_a (equ s t) = true ->
      free_list A = [n] ->
        PA_omega_theorem (lor (neg (substitution A n s)) (substitution A n t))
                         0 (ord_succ (nat_ord ((num_conn A)+(num_conn A)))).
\end{lstlisting}

\subsection{Connecting PA$_\omega$ and PA}
\label{sec:PA_equiv}

With this additional machinery in place, we can now show the embedding of the Hilbert calculus for PA into our sequent calculus for PA$_\omega$. Our first order of business is to strengthen the construction of PA by including parameters for degree and ordinal height, much like we did with PA$_\omega$. This may seem odd, since these were properties of the calculus PA$_\omega$ and without cut the degree does not seem relevant, but it is the key to allowing for the equivalence between Peano arithmetic and PA$_\omega$ to be proven. This is because the heights and degrees that will be assigned are the values of there equivalent construction within PA$_\omega$ allowing us to replace Theorem \ref{theorem:PA_equiv_bad} with Theorem \ref{theorem:PA_equiv_good}.

\begin{theorem}
\label{theorem:PA_equiv_bad}
For any closed formula $A$, if $A$ is a theorem of Peano Arithmetic, there exists some ordinal $\alpha$ and degree $d$ such that $A$ is derivable in PA$_\omega$ with a height $\alpha$ and a degree $d$.
\end{theorem}

\begin{lstlisting}[language=Coq,label=list:equivbad,abovecaptionskip=-\medskipamount]
Theorem Peano_PA_Omega_Equiv_Bad :
    forall (A : formula),
        (closed A = true) ->
            Peano_Theorem_Bad A ->
                { d : nat { alpha : ord & PA_Omega_Theorem A d alpha}}.
\end{lstlisting}

\begin{theorem}
\label{theorem:PA_equiv_good}
For any closed formula $A$, if $A$ is a theorem of Peano Arithmetic with height $\alpha$ and degree $d$, then $A$ is derivable in PA$_\omega$ with a height of $\alpha$ and a degree of $d$.
\end{theorem}

\begin{lstlisting}[language=Coq,label=list:equivgood,abovecaptionskip=-\medskipamount]
Theorem Peano_PA_Omega_Equiv :
    forall (A : formula) (d :nat) (alpha : ord),
        (closed A = true) ->
            Peano_Theorem A d alpha ->
                PA_Omega_Theorem A d alpha.
\end{lstlisting}

By removing the existential statement from the theorem, the constructive proof becomes much more straightforward, and since the steps to derive each axiom from its premises are known, these values are calculable at each step and so we can build them into the construction. An additional simplification can be made by replacing the arithmetic axioms with their unquantified forms, since by definition, Universal Generalisation can recover the original equation from that form. There is one further change we need to make from the naive translation, which is to add a trivial implication into the definition of Universal Generalisation. It is easy to prove that adding this additional constraint does not change the behaviour of the rule over closed formulae, however it does modify to inductive hypothesis in a way which is beneficial in later proofs. The proof is just a simple structural equivalence and so will be omitted here for brevity.

\begin{definition}[Change To Universal Generalisation]
\label{definition:IUG}
  We add in a requirement for closed substitutions to this definition to simplify using the rule inductively. This extra condition requires that if we want to apply universal generalisation to a theorem which is open , then the closure of that formula using an arbitrary term must also be a valid theorem in PA.
\end{definition}
\begin{lstlisting} [language=Coq,label=list:UG_Coq,abovecaptionskip=-\medskipamount]
| UG : forall (A : formula) (n m : nat) (alpha : ord),
    Peano_Theorem_Base A m alpha ->
        Peano_Theorem_Base (univ n A) m (cons alpha 0 Zero) 

| I_UG : forall (A : formula) (n m : nat) (alpha : ord),
    Peano_Theorem_Implication A m alpha ->
        (forall t, closed_t t = true ->
            Peano_Theorem_Implication (substitution A n t) m alpha) ->
        Peano_Theorem_Implication (univ n A) m (cons alpha 0 Zero)
\end{lstlisting}

\section{Gentzen's Proof Process}

With these fundamentals defined, we can now move on to the main body of the proof, which centres around a cut elimination argument.
The cut elimination argument proceeds by a triple induction. Firstly we induct over the degree of the proof. As such if we can show that given a derivation of some formula $A$ with degree $d + 1$, if it can be reduced to a derivation with degree $d$, then we can reduce it to degree 0 (necessitating that there are no cuts since any derivation with a cut has positive degree). We show this reduction can be performed by strongly inducting over the height of the derivation\footnote{This is why me must strengthen PRA with induction over the ordinals.}. With the strong inductive hypothesis for the height in place, we now perform our final induction, a structural induction on the proof tree. If the last rule applied was a short rule, then apply the structural inductive hypothesis to the premise and then reapply applying that same rule is enough to close out the proof. For any of the tall rules, the inductive hypothesis on the height will necessary apply to all premises, and so after applying that to all premises we can reapply the same tall rule to recover our endsequent. The fact that the $\omega$-rule has a side condition on the height of its premises is why the height increase caused by the cut elimination procedure must be kept track of, as we need to bound it below some ordinal less than $\epsilon_0$ to give the result of applying the $\omega$-rule a new height. Finally if the last rule that was applied is a cut we need to directly reduce the degree of the cut, by reducing the number of logical connectives in the formula being cut. This will necessarily involve do something with the tall rules, as these are the only way to introduce logical symbols. If it were instead possible to perform the cut on the premise of the tall rules rather than their conclusions this would be sufficient to reduce the degree. This would not have any effect on the conclusion of the derivation, since the only effect the cut formula has on the derivation is its degree. Therefore, we can do a case analysis on the structure of the cut formula, to figure out which deductive rules are involved before performing an inversion to perform this substitution.

\subsection{Inverting Deductive Rules}

The inversion of a deductive rule is a logically consistent substitution schema that we can define over proof trees, which allows for us to take an existing derivation, and create a new derivation in which one of the steps was omitted, and then propagate that change down through the rest of the tree. For example, if the negation rule was applied, then at some point a formula was transformed from $A$ to $\neg \neg A$, and assuming it was not cut, this means there will be a $\neg \neg A$ in the endsequent. However, we can actually go one step further, if $\neg \neg A$ appears in the endsequent, there are only two options, since only the negation and weakening rules could introduce this subformula. This gives us enough information to show that if $\neg \neg A$ appears in an endsequent, the same formula with $\neg \neg A$ replaced with $A$ must have a well-formed derivation, since if weakening introduced $\neg \neg A$ it could just as easily have introduced $A$ instead, and if it was introduced with the negation rule, simply not applying the rule at that step would be sufficient.

\begin{example}[Negation Inversion Process]
If the endsequent has a formula of the form $(C \vee \neg \neg A) \vee D$, then we know that the derivation must be a well-formed proof tree with one of these two substructures, where $\neg \neg A$ was not principle in any of the $D_i$ that are tall rules.

\begin{tabular}{cc}
\AxiomC{$\vdots$}
\UnaryInfC{$A \vee E$}
\RightLabel{Negation}
\UnaryInfC{$\neg \neg A \vee E$}
\UnaryInfC{$\vdots$}
\RightLabel{$D_1, \cdots, D_n$}
\UnaryInfC{$(C \vee \neg \neg A) \vee A$}
\DisplayProof
&
\AxiomC{$\vdots$}
\UnaryInfC{$E$}
\RightLabel{Weakening}
\UnaryInfC{$\neg \neg A \vee E$}
\UnaryInfC{$\vdots$}
\RightLabel{$D_1, \cdots, D_n$}
\UnaryInfC{$(C \vee \neg \neg A) \vee A$}
\DisplayProof
\end{tabular}

\

In the left case, if we do not apply the inversion at this step, then the we can continue applying all the $D_i$ to the premise instead.

\AxiomC{$\vdots$}
\UnaryInfC{$A \vee E$}
\UnaryInfC{$\vdots$}
\RightLabel{$D_1, \cdots, D_n$}
\UnaryInfC{$(C \vee \neg \neg A) \vee A$}
\DisplayProof

\

In the right case, we instead change the weakening to add $A$ rather than $\neg \neg A$, before continuing to apply all the $D_i$ in order.

\AxiomC{$\vdots$}
\UnaryInfC{$E$}
\RightLabel{Weakening}
\UnaryInfC{$A \vee E$}
\UnaryInfC{$\vdots$}
\RightLabel{$D_1, \cdots, D_n$}
\UnaryInfC{$(C \vee \neg \neg A) \vee A$}
\DisplayProof

In either case for all of the $D_i$ that are not instances of contraction with $\neg \neg A$ principle, they will still apply with their parameters adjusted to account for the change from $\neg \neg A$ to $A$. If one of the $D_i$ is a contraction with $\neg \neg A$ principle then we must start this process again from the now modified premise of the contraction and invert this other negated formula first before continuing with the contraction and following $D_i$.
\end{example}

\begin{lstlisting}[language=Coq,label=list:neg_invCoq,abovecaptionskip=-\medskipamount]
Fixpoint dub_neg_sub_ptree_fit  (P : ptree) (E : formula) (S : subst_ind) : ptree :=
match P, S with
| node A, _ => P

| exchange_ab A B d alpha P', lor_ind S_B S_A =>
    exchange_ab (dub_neg_sub_formula A E S_A) (dub_neg_sub_formula B E S_B)
      d alpha (dub_neg_sub_ptree_fit P' E (lor_ind S_A S_B))

(* Other short rules omitted for brevity *)

| contraction_ad A D d alpha P', lor_ind S_A S_D =>
    contraction_ad (dub_neg_sub_formula A E S_A) (dub_neg_sub_formula D E S_D)
      d alpha (dub_neg_sub_ptree_fit P' E (lor_ind (lor_ind S_A S_A) S_D))

| weakening_ad A D d alpha P', lor_ind S_A S_D =>
    (weakening_ad (dub_neg_sub_formula A E S_A) (dub_neg_sub_formula D E S_D)
      d alpha (dub_neg_sub_ptree_fit P' E S_D))

| demorgan_ab A B d1 d2 alpha1 alpha2 P1 P2, _ => P

| demorgan_abd A B D d1 d2 alpha1 alpha2 P1 P2, lor_ind S_AB S_D =>
    demorgan_abd A B (dub_neg_sub_formula D E S_D) d1 d2 alpha1 alpha2
      (dub_neg_sub_ptree_fit P1 E (lor_ind (0) S_D))
      (dub_neg_sub_ptree_fit P2 E (lor_ind (0) S_D))

| negation_a A d alpha P', _ =>
    (match eq_f A E, S with
    | true, (1) => ord_up (ord_succ alpha) P'
    | _, _ => P
    end)

| negation_ad A D d alpha P', lor_ind S_A S_D =>
    (match eq_f A E, S_A with
    | true, (1) => ord_up
                      (ord_succ alpha)
                      (dub_neg_sub_ptree_fit P' E (lor_ind (non_target A) S_D))
    | _, _ =>  negation_ad A (dub_neg_sub_formula D E S_D)
                  d alpha (dub_neg_sub_ptree_fit P' E (lor_ind (non_target A) S_D))
    end)

(* Other rules omitted for brevity *)
end.
\end{lstlisting}

To maintain the structure of the derivation, we will still need to increase the height by 1 in the case we we instead skip applying the negation, so that this process only changes the formulae rather than the height or degree. With that assurance, we can now using the substitution indicators from before, we can now define inversion principles for the negation, demorgan and $\omega$ deductive rules and, and prove that the resultant proof trees are well-formed.

\begin{theorem}[Inversion of Negation, DeMorgan, and $\omega$]
\

If there exists a well-formed proof tree with endsequent $(C \vee \neg\neg A) \vee D$, there is a well-formed proof tree with endsequent $(C \vee A) \vee D$.

If there exists a well-formed proof tree with endsequent $(C \vee \neg (A \vee B)) \vee D$, there are a well-formed proof trees with endsequent $(C \vee \neg A) \vee D$ and $(C \vee \neg B) \vee D$.

If there exists a well-formed proof tree with endsequent $(C \vee \forall x_n, A(x_n)) \vee D$, there is a well-formed proof tree with endsequent $(C \vee A(n)) \vee D$ for each $n \in \mathbb{N}$.
\end{theorem}

\begin{lstlisting}[language=Coq,label=list:inv_wfCoq,abovecaptionskip=-\medskipamount]
Lemma dub_neg_wf :
    forall (P : ptree) (E : formula),
        well_formed P ->
            forall (S : subst_ind),
                subst_ind_fit (ptree_formula P) S = true ->
                    well_formed (dub_neg_sub_ptree P E S).

Lemma demorgan_wf_1 :
    forall (P : ptree) (E : formula),
        well_formed P ->
            forall (S : subst_ind),
                subst_ind_fit (ptree_formula P) S = true ->
                    well_formed (demorgan1_sub_ptree P E S).

Lemma demorgan_wf_2 :
    forall (P : ptree) (E : formula),
        well_formed P ->
            forall (S : subst_ind),
                subst_ind_fit (ptree_formula P) S = true ->
                    well_formed (demorgan2_sub_ptree P E S).

Lemma omega_wf :
    forall (P : ptree) (E : formula),
        well_formed P ->
            forall (S : subst_ind),
                subst_ind_fit (ptree_formula P) S = true ->
                    (forall (n : nat), 
                        well_formed (omega_sub_ptree P E S n)).
\end{lstlisting}

\

Due to a subtlety of the quantification rule, this technique does not work for the final tall rule. The issue with inverting the quantification rule is that since this is a classical system, $\neg \forall$ is the existential quantifier, and so applying this rule in a constructive framework like Coq actually loses information. The premise has a specific counter example ($\neg A[x_i/n]$) while the conclusion contains the weaker statement ($\neg \forall n, A[x_i/n]$). So the question becomes, when faced with an instance of $\neg \forall n, A[x_i/n]$ in the end sequent, what do we replace it with. If the formula was introduced with weakening it is obviously unimportant, and similarly if there is only a single instance of quantification which lead to the formula then we know exactly what to replace it with, however the contraction rule means that there could be two quantifications in the ancestry of the formula, which had different values of $n$.

\begin{example}[Incompatibility of Contraction With Naively Inverted Quantification]
\begin{tabular}{ccc}
\AxiomC{$\vdots$}
\UnaryInfC{$\neg A[x_i/n_1] \vee \neg A[x_i/n_2]$}
\RightLabel {Quantification 1}
\UnaryInfC{$\neg(\forall t, A[x_i/t]) \vee \neg A[x_i/n_2]$}
\RightLabel {$E_1$}
\UnaryInfC{$\neg A[x_i/n_2] \vee \neg(\forall t, A[x_i/t])$}
\RightLabel {Quantification 1}
\UnaryInfC{$\neg(\forall t, A[x_i/t]) \vee \neg(\forall t, A[x_i/t])$}
\RightLabel {Contraction 1}
\UnaryInfC{$\neg(\forall t, A[x_i/t])$}
\DisplayProof

&

\scalebox{1.5}{$\Longrightarrow$}

&

\AxiomC{$\vdots$}
\UnaryInfC{$\neg A[x_i/n_1] \vee \neg A[x_i/n_2]$}
\RightLabel {$E_1$}
\UnaryInfC{$\neg A[x_i/n_2]  \vee \neg A[x_i/n_1]$}
\RightLabel {Contraction 1}
\UnaryInfC{$\neg A[x_i/n_2] \vee D$}
\DisplayProof
\end{tabular}

\

This final contraction in the transformed derivation is obviously nonsense, unless $n_1$ and $n_2$ happen to be equal, but this need not be the case.
\end{example}

While a potential avenue for solving this issue would be to first transform the derivation into a form which derives all instances of quantification for a given formula $A$ used the same value of $n$, the machinery that we have covered thus far is insufficient to describe the necessary transformation, and would certainly not be able to prove that such an operation was permissible. Luckily, there is a much more simple solution which comes from G\"{o}del's\cite{godel} alternative proof for Gentzen's result.

\subsection{G\"{o}del's Method}
\label{sec:godel}

What G\"{o}del came up with during his review was an elegant solution to the problem with inverting quantification by forgoing this idea of a global counter-example entirely. He discovered that instead you could instead take a different approach to the cut elimination process itself when dealing with universally quantified cuts. In doing so he found a solution that was much more amenable to the constructive nature of a formalised proof, since it avoided asserting classicality and solved the problem of finding the ``correct'' counterexample since quantification was never inverted. While this discovery was a great help, we were required to make some further modifications to the argument in order to make this approach correctly interface with the sequents in our formalisation.

The key idea behind G\"{o}del's approach is realising that the existence of something of the form $\neg \forall x, A(x)$ in the endsequent tells us that after the weakening or quantification which introduced the subformula, it was not principle in any of the following tall rules or cuts. It was always a side formula in such a deductive step, it could only be principle in a short rule. As such, as far as the structure of the sequent is concerned the specifics of subformula are irrelevant and it could be replaced by some other subformula and the derivation would progress in the same way. As such, if you have a proof tree with an instance of quantification whose principle formula will be cut, you can replace that instance of quantification with an instance of cut (using the inversion of the $\forall$ as the other premise of the cut) and you would still arrive at the same endsequent. Moreover, is is not difficult to show that this transformation will produce a well-formed proof tree.

\

\begin{example}[G\"{o}del's Transformation]
Given the following derivation of $C$ ending in a cut, we can create a new derivation of $C$ with smaller cuts.

\AxiomC{$\vdots$}
\UnaryInfC{$C \vee (\forall t, A[x_n/t])$}

\AxiomC{$\vdots$}
\UnaryInfC{$\neg(A[x_i/n_1]) \vee \neg(A[x_i/n_2])$}
\RightLabel{Quantification 2}
\UnaryInfC{$\neg(\forall t, A[x_i/t]) \vee \neg(A[x_i/n_2])$}
\RightLabel{$E_1$}
\UnaryInfC{$\neg A[x_i/n_2]\vee \neg(\forall t, A[x_i/t])$}
\RightLabel {Quantification 2}
\UnaryInfC{$\neg(\forall t, A[x_i/t]) \vee \neg(\forall t, A[x_i/t])$}
\RightLabel {Contraction 1}
\UnaryInfC{$\neg(\forall t, A[x_i/t])$}
\RightLabel {Cut 1}
\BinaryInfC{$C$}
\DisplayProof

\vspace*{1cm}

First we invert the $\omega$-rule used in the derivation of $C \vee (\forall t, A[x_n/t])$ twice, once with the value of $n_1$, ($P_1$) and once with the value of $n_2$, ($P_2$). Then we replace each instance of quantification of the value $n_i$ by a cut with $P_i$.

\AxiomC{$P_2$}
\UnaryInfC{$C \vee A[x_i/n_2]$}

\AxiomC{$P_1$}
\UnaryInfC{$C \vee A[x_i/n_1]$}

\AxiomC{$\vdots$}
\UnaryInfC{$\neg(A[x_i/n_1]) \vee \neg(A[x_i/n_2])$}
\RightLabel {Cut 1}
\BinaryInfC{$C\vee\neg(A[x_i/n_2])$}
\RightLabel{$E_1$}
\UnaryInfC{$\neg A[x_i/n_2]\vee C$}
\RightLabel {Cut 1}
\BinaryInfC{$C \vee C$}
\RightLabel {Contraction 1}
\UnaryInfC{$C$}
\DisplayProof
\end{example}

\

\

Unfortunately, this idea of $C$ being present is central to preserving the structure of the proof tree, and it is not always the case that there will be a side formula $C$, sometimes the only side formula will be $D$, on the negated side of the cut. In this case, the structure of the formula changes and any further steps may become invalid. Exchange is especially impacted, since the effect of an exchange is entirely dependant on the structure of the formula.

\begin{example}[Bad G\"{o}del Transformation]

\

If one were to try and perform the G\"{o}del transformation on a cut without a left side formula you would see that the derivation:

\AxiomC{$\vdots$}
\UnaryInfC{$\forall t, A[x_n/t]$}

\AxiomC{$\vdots$}
\UnaryInfC{$\neg A[x_n/m] \vee (D_1 \vee D_2)$}
\RightLabel{Quantification 2}
\UnaryInfC{$\neg (\forall t, A[x_n/t]) \vee (D_1 \vee D_2)$}
\RightLabel{$E_1$}
\UnaryInfC{$(D_1 \vee D_2) \vee \neg (\forall t, A[x_n/t])$}
\RightLabel{Negation 2}
\UnaryInfC{$\neg \neg (D_1 \vee D_2) \vee \neg (\forall t, A[x_n/t])$}
\RightLabel{$E_1$}
\UnaryInfC{$\neg (\forall t, A[x_n/t]) \vee \neg \neg (D_1 \vee D_2)$}

\RightLabel{Cut 2}
\BinaryInfC{$\neg \neg (D_1 \vee D_2)$}
\DisplayProof

\

Would become:

\AxiomC{$\vdots$}
\UnaryInfC{$A[x_n/m]$}

\AxiomC{$\vdots$}
\UnaryInfC{$\neg A[x_n/m] \vee (D_1 \vee D_2)$}
\RightLabel{Cut 2 $(A[x_n/m]$}
\BinaryInfC{$D_1 \vee D_2$}
\RightLabel{$E_1$}
\UnaryInfC{$D_2 \vee D_1$}
\RightLabel{Negation 2}
\UnaryInfC{$\neg \neg D_2 \vee D_1$}
\RightLabel{$E_1$}
\UnaryInfC{$D_1 \vee \neg \neg D_2$}
\DisplayProof

\

Which no longer has the same endsequent.
\end{example}

\

The solution to this problem is to modify the behaviour when we are the case of replacing a \lstinline{cut_ad}. After inverting the $\omega$-rule to obtain a value of $A[x_i/n]$, we weaken by the side formula we do have, $D$ and now we can do a \lstinline{cut_cad} instead. When we propagate this change down, at the point where the original cut would have occurred, we are left with a formula of the form $D \vee D$ and so a single contraction is sufficient to provide a derivation of $D$ with reduced the degree. Applying the weakening at this earlier point raises some interesting difficulties when managing the height of the new derivation. As part of the proof we are trying to show that if the original derivation has height $\alpha$ then the derivation with lower degree has height at most $2^\alpha$. By adding this weakening, we increase the height earlier in the process, meaning that the compounding effects on height that the later steps involving cut have can violate the bound.\footnote{This is actually a very specific edge case, requiring that both premises of the original cut were finite and equal, all other instances preserve the bound.} This was not covered in G\"{o}del's paper, as his system instead used implicit exchange and contraction rules, meaning that these intermediate short rules were not present in either form. Since short rules do not impact the height, G\"{o}del's approach provides a complete solution. However, implementing a sequent calculus which implicitly uses exchange and contraction would have involved rewriting the majority of the code, so we instead found a ``native'' solution to the issue that fit within the existing framework. We instead tackled the problem with the bounds directly by introducing that extra case to the definition of the height of a proof tree. This is why \lstinline{cut_ad} increases the height by 2, because that mean it has the same effect on height as applying both \lstinline{weakening_ad} and \lstinline{cut_cad}. With this fix in place we have now completed the cut elimination algorithm for the universally quantified formulae.

\

\subsection{Cut Elimination}
With our modified G\"{o}del's method in mind for the universally quantified case we can now return to proving cut elimination in full.

\begin{theorem}[Cut Elimination]
\label{theorem:cut_elim}
For any formula $A$, if there is a well-formed proof tree with endsequent $A$ and degree $d$ and height $\alpha$, then there exists another well-formed proof tree with endsequent $A$ and degree 0 with unknown height.
\end{theorem}

\begin{lstlisting}[language=Coq,label=list:cut_elim_full_Coq,abovecaptionskip=-\medskipamount]
Theorem cut_elim :
    forall (P : ptree) (A : formula) (d : nat) (alpha : ord),
        (well_formed P) * (ptree_formula P = A) *
        (ptree_deg P = d) * (ptree_ord P = alpha) ->
            {beta : ord & provable A 0 beta}.
\end{lstlisting}

To prove Theorem \ref{theorem:cut_elim} we will induct over the degree of the proof, giving the following proof state.

\begin{definition}[Outermost Induction - Degree]
Assume that for any proof tree $P$ if the degree of $P$ is $d$, then there exists some other derivation $P^\prime$ with the same endsequent as $P$ and degree 0 and unknown height.
Show that for some proof tree $Q$ with degree $d+1$, there exists a derivation $Q^\prime$ with the same endsequent as $Q$ and degree 0.
\end{definition}

\begin{lstlisting}[language=Coq,label=list:ind_hyp_1Coq,abovecaptionskip=-\medskipamount]
d : nat
IHd : forall (P : ptree) (B : formula) (beta : ord),
        (well_formed P) * (ptree_formula P = B) *
        (ptree_deg P = d) * (ptree_ord P = beta) ->
            {gamma : ord & provable B 0 gamma} 
Q : ptree
A : formula
alpha : ord
Q_wf : well_founded Q
Q_formula : ptree_formula Q = A
Q_deg : ptree_deg Q = S d
Q_height : ptree_ord Q = alpha

------------------------------------

$\vdash$ {delta : ord & provable A 0 delta}.

\end{lstlisting}

Therefore, in order to close out the argument with the inductive hypothesis, it is sufficient to define a transformation from $Q$ to some well-formed tree $P$ with degree $d$ and the same endsequent. To perform this reduction of the degree by a single step, we will prove an auxiliary theorem.

\begin{theorem}[Auxillary Cut Elimination Result]
\label{theorem:cut_aux}
For any ordinal $\alpha$ in normal form, for all well-formed proof trees with degree $d+1$ and height $\alpha$, there exists a well formed proof tree with the same endsequent, degree $d$ and height $2^\alpha$.
\end{theorem}

\begin{lstlisting}[language=Coq,label=list:cut_elim_aux_Coq,abovecaptionskip=-\medskipamount]
Theorem cut_elim_aux :
    forall (alpha : ord), nf alpha ->
       forall (P : ptree) (A : formula) (d : nat),
          (well_formed P) * (ptree_formula P = A) *
          (ptree_deg P = S d) * (ptree_ord P = alpha) ->
              provable A d (ord_2_exp alpha).
\end{lstlisting}

\

The proof of the auxiliary result comes through a strong induction on the height of the proof tree. There will be two base cases for this induction, since certain inequalities relating exponentiation and addition, only hold for values of at least 2. For height 0 the reduction is trivial, since cut is a height increasing operation, a height 0 derivation can not contain a cut at all, so could not have positive degree in the first place. A height 1 derivation also can not involve cut, since a height 1 derivation involving cut must by definition have two height 0 premises\footnote{Or not be possible at all in the case of \lstinline{cut_ad}, by the adjustment in the previous section.}. However, it is also true that at least one of the premises of a cut must contain a disjunction and no height 0 derivations contain a disjunction. As such both base cases are dealt with.

\

To complete the proof in the other cases, we must now induct over the structure of the proof tree $P$. This will give us both the last deductive rule which was applied ($\mathcal{D}$), as well as list of premises $\{P_i\}$, with heights $\alpha_i$, endsequents $A_i$ and degrees $d_i$.

\

If $\mathcal{D}$ is a short rule, then the height and degree of the premise are the same as the height and degree of the conclusion. As such the structural inductive hypothesis provides a map from the premise $P_0$ to a new proof tree $Q$ with one lower degree and height $2^\alpha$ and the same endsequent as $P_0$. Reapplying $\mathcal{D}$ to $Q$ will again not change the height or degree, and allow us to recover the original endsequent.

\

If instead $\mathcal{D}$ was a tall rule, we know the height of all the $P_i$ are lower than the height of $P$. As such the strong inductive hypothesis will apply to all of them, decreasing their degree by 1, and changing their heights from $\alpha_i$ to $2^{\alpha_i}$, while preserving the endsequents $A_i$. Again we can now reapply $\mathcal{D}$ to regain the original endsequent at the lower degree, as the degree of a tall rule is the maximum of its premises, the uniform reduction in the degrees of the premises decreases the final degree. Finally, since the height of a tall rule is one greater than the maximum (or supremum in the case of the $\omega$-rule)s of the heights of the premises for that rule, the inequality $2^\alpha + 1 < 2^{\alpha + 1}$ which holds for all $1 < \alpha$ shows that the bound on heights is preserved.

Now the only rules that need to be dealt with are the 3 types of cut. 

\begin{example}[The 3 Forms of Cut]

\

\

\normalfont

\begin{tabular}{ccc}
\AxiomC{$C \vee A$}
\AxiomC{\hspace*{-2mm}$\neg A$}
\RightLabel{cut\_ca}
\BinaryInfC{$C$}
\DisplayProof
&
\AxiomC{$A$}
\AxiomC{\hspace*{-3mm}$\neg A \vee D$}
\RightLabel{cut\_ad}
\BinaryInfC{$D$}
\DisplayProof
&
\AxiomC{$C \vee A$}
\AxiomC{$\neg A \vee D$}
\RightLabel{cut\_cad}
\BinaryInfC{$C \vee D$}
\DisplayProof
\\
\end{tabular}

\end{example}

\

For each of these rules the first step is the same as for the tall rule, we apply the inductive hypothesis on height to $P_1$ and $P_2$ to generate new proof trees with lower degree, $Q_1$ and $Q_2$. This decreasing of the degree of the premises is necessary because sometimes the lowest cut, which is the cut which will be affected first by a structural induction, is not the cut of highest degree in the proof tree, and since the degree of a cut is the maximum of the degrees of its premise and the number of logical symbols in $\neg A$, we need to reduce the premises to be sure a reduction will be achieved. We can then move on to decreasing the size of the cut formula $A$. This is done by breaking into case for the different structures of $A$, is it an atomic formula, negated, disjunctive or universally quantified. The universally quantified case has been dealt with for all three types of cut in the previous section.

\

If the formula being cut is atomic, then since it is closed formula, by definition exactly one of $A$ and $\neg A$ is an axiom of PA$_\omega$. As such which ever subformula is not an axiom must have been introduced through weakening. If $\neg A$ was the axiom, the same process will work for all three types of cut. In this case, there exists some point in the derivation $Q_1$ where $A$ was added through an instance of weakening. At this point we instead introduce $D$ with the weakening, and then perform the rest of the deductive steps of $P_1$, which will result in a derivation of the endsequent which did not have to cut $A$, and will have the same height as $Q_1$. If $A$ was instead the axiom, then instead there was a point in $Q_2$ where $\neg A$ was introduced by weakening. If $\mathcal{D}$ was \lstinline{cut_ca} or \lstinline{cut_cad}, then we replace the weakening of $\neg A$ by a weakening with $C$ and we end with a derivation of of the endsequent without a cutting $A$ and with the same height as $Q_1$. If the deductive rule was instead $\lstinline{cut_ad}$, then we can not replace it with $C$ since there is no $C$, we will instead replace the weakening of $\neg A$ with a weakening by $D$. This will result in a proof tree with endsequent $D \vee D$, with the same height as $Q_2$. We can then apply a contraction, to recover a derivation of $D$ of the correct form, as the contraction will not further increase the height. 

\

For the remaining two cases, the proof is the same for all three forms of cut, so we shall only show the \lstinline{cut_cad} case.

If the cut formula is negated, then $Q_1$ has $\neg A$ as a subformula, and $Q_2$ has $\neg \neg A$. By performing an inversion of negation on $Q_2$, we can transform $Q_2$ into some $Q_2^\prime$ which has $A$ as a subformula instead of $\neg \neg A$. After applying an exchange to $Q_1$ and $Q_2^\prime$, we now have the premises for a cut of $A$, rather than $\neg A$, and so the degree will be one lower, because the inversion process preserves height and degree, and the only other rules needed are exchange, which also has no effect on these values. Also, by the same inequality that was used for the tall rules, we have preserved the bound on the height.

Finally, if the cut formula is disjunctive, we will have to use the both inversions of the Demorgan rule on $\neg (A_1 \vee A_2)$ on $Q_2$ to get proof trees and $Q_{2A_1}$ and $Q_{2A_2}$ which have replaced the negated disjunction with the negated left and right disjuncts respectively. We can then perform a cut of $A_1$ with $Q_1$ and $Q_{2A_1}$\footnote{This is on modulo applying some of the associativity results from earlier to bracket the formulae, but these results did not change degree or height.}. This resultant proof tree can then be used in a cut of $A_2$ along with $Q_{2A_2}$. After performing some more exchanges and contractions to remove the extra copy of $D$ that this added, we have now been left with same endsequent as the original conclusion of $\mathcal{D}$, by performing two smaller cuts. This leaves us with a resultant degree equal to the maximum of: the degree of $Q_1$, the degree of $Q_{2A1}$, the degree of $Q_{2A_2}$, the size of $\neg A_1$ and the size of $\neg A_2$. Since inversions and exchange do not change the height or degree, we know that all of the $Q$ premises have lower degree than the original premises $P_1$ and $P_2$. Meanwhile, for the cut formula components the the inequality $\max\{a+1,b+1\} < a+b+2$ is sufficient to show a reduction occurred, and this holds for all values of $a$ and $b$ in $\mathbb{N}$. However, since there are now two cuts, (along with many exchanges and contractions), the height is potentially increased by 2, depending on the relative heights of the premises. Therefore, it is necessary that the inequality $2^\alpha + 2 < 2^{\alpha + 1}$ holds. It is for this reason that we required the secondary base case, as this is true for all $1 < \alpha$.

\

With this, we have now shown that for any derivation with a given degree and height, a derivation with the same endsequent and degree 1 lower can be constructed within the bound on height, which completes the proof of Theorem \ref{theorem:cut_aux}. We can now use Theorem \ref{theorem:cut_aux}, to close out the inductive step of Theorem \ref{theorem:cut_elim}, proving cut elimination holds for PA$_\omega$.

\

\subsection{Consistency of PA$_\omega$}

To say that a proof calculus is consistent is to say that the calculus can not derive a contradiction, or equivalently, that false statements can not be derived. This is a highly desirable trait for a calculus to have, since inconsistent systems will by definition produce unreliable results.

With the cut elimination argument completed, the proof for the consistency of PA$_\omega$ is relatively straight forward. Following Gentzen, we simply show that there is no proof of some given false statement which has degree 0. False statements have many forms, but the simplest method is to consider disjunctions of some equality which does not hold, for example (0=1). We call these formulae dangerous disjuncts, and if we can show that all such dangerous disjunctions do not have a degree 0 derivation, then there can not be any derivation of these formulae and thus they are not a theorem in the logic. 

\begin{lemma}
If a proof tree $P$ proves the disjunction of a dangerous formula, then $P$ must have non-0 degree.
\end{lemma}
\begin{lstlisting}[language=Coq,label=list:danger_nonzero,abovecaptionskip=-\medskipamount]
Lemma danger_not_deg_0 :
    forall (P : ptree) (A : formula) (d : nat) (alpha : ord),
        (well_formed P) * (ptree_formula P = A) *
        (ptree_deg P = d) * (ptree_ord P = alpha) ->
            dangerous_disjunct A = true ->
                0 < d.
\end{lstlisting}

This proof is quite straightforward, by definition (0=1) is not an initial sequent, so the derivation must involve some deductive rules. We then check each of the rules to see deductions can produce a dangerous disjunct from premises which are not dangerous. All of the short rules simply transform dangerous disjuncts into other dangerous disjuncts, so can not be the source of a dangerous disjunct. For any of the tall rules except weakening, they introduce some number of $\neg$ and $\forall$ symbols, to their principle formula, and these symbols are not present in a dangerous disjunct. Weakening introduces a new disjunction to a formula, but if all disjuncts are (0=1) after the weakening then this must have also been true before the application of weakening as well. Uses of cut can produce a dangerous disjunct from safe premises. Consider two derivations with endsequents $A \vee (0 = 1)$ and $\neg A$ as shown in the diagram below. Applying cut to these premises would yield a derivation of (0=1), but neither premise was a dangerous disjunct. However, since cut always results in a non-zero degree, all dangerous derivations must have non-zero degree. Using our cut elimination result from before, this means that there can be no derivation of a dangerous disjunct, and furthermore PA$_\omega$ must be consistent, else the inconsistency would yield a derivation of dangerous disjuncts.

\begin{lstlisting}[language=Coq,label=list:inconsistentdanger,abovecaptionskip=-\medskipamount]
Lemma inconsistent_danger :
    forall (A : formula) (n1 n2 : nat) (alpha1 alpha2 : ord),
        PA_omega_theorem A n1 alpha1 ->
            PA_omega_theorem (neg A) n2 alpha2 -> False.
\end{lstlisting}

\subsection{Equivalence of PA and PA$_\omega$}

With the consistency of PA$_\omega$ proven, we now need only show that this implies the consistency of Peano Arithmetic. We do this by proving an equivalence between closed theorems in each system. Unfortunately, the simplest approach of building a direct equivalence fails. The rule for Universal Generalisation only acts on formulae which are not closed. As such any statement similar to ``For all closed formulae $A$, if $A$ is a theorem in Peano Arithmetic then $A$ is also a theorem in PA$_\omega$.'' would fail to be applicable for universal generalisation. Instead we will actually prove a stronger equivalence, after defining something called the closure of a formula.

The closure of a formula $A$, is the result of choosing some closed term $c$ and then replacing all unbound free variables present in $A$ with $c$. For example, the closure of $x_4 + 1 = x_7 \times 3$ by 2 would be $2 + 1 = 2 \times 3$. As such, the closure of any formula will be closed, and the closure of a closed formula $A$ is just $A$ without any changes, as there are no unbound free variables to be substituted. Since the operation acts as the identity on closed formulas, we can replace $A$ with the closure of $A$ and gain the same information\footnote{This is also why the trivial implication needed to be added to the definition of universal generalisation, to account for this replacement}.

\begin{theorem}
\label{theorem:PA_equiv}
 For any formula $A$. If $A$ is a theorem of Peano Arithmetic with degree $d$ and height $\alpha$. Then for all closed terms $t$ the closure of the formula $A$ by $t$ is provable with degree $d$ and height $\alpha$.\vspace*{-2mm}
\end{theorem}
\begin{lstlisting}[language=Coq,label=list:peano_equiv_pa,abovecaptionskip=-\medskipamount]
Theorem PA_closed_PA_omega :
    forall (A : formula) (d : nat) (alpha : ord),
        Peano_Theorem A d alpha ->
            (forall t, closed_t t = true ->
             PA_omega_theorem (closure A t) d alpha).
\end{lstlisting}

When combined with definition \ref{definition:IUG}, the inductive hypothesis of theorem \ref{theorem:PA_equiv}, will now apply in the case of universal generalisation, since we only require that the closure of the known to be unclosed formula is a theorem, rather than requiring that the formula be both closed and a theorem. Since the change to the inductive hypothesis is a bit subtle, we have provided below a snapshot of the proof state at the start of the case for universal generalisation. 

\begin{example}[Proof State in the Case of Universal Generalisation]
Because the inductive definition of \lstinline{I_UG} references both a specific element, and a parameterised family of elements, it generates an inductive hypothesis for both the single element and that family. This second inductive hypothesis exactly describes the family of premises required for an application of the $\omega$-rule for the conclusion. However, without the modified definition of universal generalisation, only T1 would be generated by the induction and thus IHT2 would not exist in the proof state. 
\end{example}

\begin{lstlisting}[language=Coq,label=list:UG_proofstate,abovecaptionskip=-\medskipamount]
A : formula
n d : nat
alpha : ord
T1 : Peano_Theorem A d alpha
T2 : forall (t : term), closed_t t = true ->
        Peano_Theorem (substitution A n t) d alpha
IHT1 : forall (c : c_term), PA_omega_theorem (closure A c) d alpha
IHT2 : forall (t : term), closed_t t = true -> forall (c : c_term),
              PA_omega_theorem (closure (substitution A n t) c) d alpha

-----------------------------------------------

$\vdash$ PA_omega_theorem (closure (univ n A) c) d (ord_succ alpha)
\end{lstlisting}

With this change we have shown that PA$_\omega$ contains the deductive rule of Universal Generalisation. The other deductive rule is even simpler, as Modeus Ponens is exactly \lstinline{cut_ad} after having applied the translation from Peano Arithmetic to PA$_\omega$.

\

For the 5 First Order Logic Axioms, each can be derived by applying a certain series of deductions to some number of instance of the the Law of the Excluded Middle. Each Axioms has its own path, but the general principle involves invoking the Law for some specific formula, and then applying a series of exchanges, weakenings and the DeMorgan rule to produce the required result. The specific details of this process are not very enlightening, so we shown only one example here, however several others can be seen in the appendices.

\begin{example}[Derivation of FOL1 in PA$_\omega$]

\

We wish to derive the 1st Axiom of First Order Logic: $A \to (B \to A)$.

Which in our representation is the formula: $\neg A \vee (\neg B \vee A)$.

\normalfont

\RightLabel{LEM (A)}
\LeftLabel{\scriptsize$0, \alpha$}
\AxiomC{$\vdots$}
\UnaryInfC{$\neg A \vee A$}
\LeftLabel{\scriptsize$0,\alpha+1$}
\RightLabel{Weakening}
\UnaryInfC{$\neg B \vee (\neg A \vee A)$}
\LeftLabel{\scriptsize$0,\alpha+1$}
\RightLabel{Exchange 1}
\UnaryInfC{$(\neg A \vee A) \vee \neg B$}
\LeftLabel{\scriptsize$0,\alpha+1$}
\RightLabel{Exchange 2}
\UnaryInfC{$(\neg A \vee \neg B) \vee A$}
\LeftLabel{\scriptsize$0,\alpha+1$}
\RightLabel{Associativity 1}
\UnaryInfC{$\neg A \vee (\neg B \vee A)$}
\DisplayProof
\end{example}

\

For the arithmetic axioms, we will actually prove more general versions of them, since it will simplify the proofs greatly. This generalisation will come from lifting the universal quantifier from the equation ($\forall$), out of the language into a Coq forall. We can do this because we have shown that Universal Generalisation is admissible in the calculus, and it is the inverse of this operation. So we will be able to recover our original axioms from these more general ones.

\begin{example}[Requantified Arithmetic Axioms]
We show how the statement of transitivity of equality changes from the original formulation to the more general version.

\normalfont

\begin{lstlisting}[language=Coq,label=list:a,abovecaptionskip=-\medskipamount]
eq_trans_old :=
forall (i j k : nat),
    univ i (
        univ j (
            univ k (
                lor (neg (atom (equ (var i) (var j))))
                    (lor (neg (atom (equ (var j) (var k))))
                         (atom (equ (var i) (var k)))))))
\end{lstlisting}

\begin{lstlisting}[language=Coq,label=list:a,abovecaptionskip=-\medskipamount]
eq_trans_new :=
forall (t s r : term),
    lor (neg (atom (equ t s)))
        (lor (neg (atom (equ s r)))
              (atom (equ t r)))
\end{lstlisting}

\

And if we set $t = \mathrm{var \ } i, s = \mathrm{var \ } j, r = \mathrm{var \ } k$, and apply universal generalisation this would recover the original formula from the new formula.
\end{example}

Deriving the transitivity of equality axiom is simple. We just weaken a specific instance of the Law of the Excluded Middle.

The two axioms relating equality to the successor function are just weakening applied to an initial sequent of PA$_\omega$.

In all 3 cases, what value must be added through weakening depends on if the equality holds or not.

\begin{example}[Non-Trivial Derivations of the Arithemtic Axioms of Peano Arithmetic]

\

We define $t \approx s$ to mean that the closure of $t$ and $s$ have the same evaluation.

\

We define $t == s$ as shorthand for \lstinline{atom (equ t s))}.

\

\normalfont
\begin{tabular}{c}
a) Derivation of Transitivity of Equality when $t_0 \approx t_1$\vspace*{-4mm}\\
\\
\AxiomC{$\vdots$}
\LeftLabel{\scriptsize$0,\alpha$}
\RightLabel{Theorem \ref{theorem:ELEM}}
\UnaryInfC{$\neg (t_1 == t_2) \vee (t_0 == t_2)$}
\LeftLabel{\scriptsize$0,\alpha+1$}
\RightLabel{Weakening}
\UnaryInfC{$\neg (t_0 == t_1)  \vee (\neg (t_1 == t_2) \vee (t_0 == t_2))$}
\DisplayProof
\\
\\
b) Derivation of Transitivity of Equality when $t_0 \not\approx t_1$\vspace*{-2mm}\\
\\
\AxiomC{$\neg (t_0 == t_1)$}
\LeftLabel{\scriptsize$0,1$}
\RightLabel{Weakening}
\UnaryInfC{$(\neg (t_1 == t_2) \vee (t_0 == t_2)) \vee \neg (t_0 == t_1)$}
\LeftLabel{\scriptsize$0,1$}
\RightLabel{$E_1$}
\UnaryInfC{$\neg (t_0 == t_1)  \vee (\neg (t_1 == t_2) \vee (t_0 == t_2))$}
\LeftLabel{\scriptsize$0,\alpha+1$}
\RightLabel{Increase Height}
\UnaryInfC{$\neg (t_0 == t_1)  \vee (\neg (t_1 == t_2) \vee (t_0 == t_2))$}
\DisplayProof

\\
\\
c) Derivation of Equality Implying Successor Equality when $t_0 \approx t1$\vspace*{-2mm}\\
\\
We know that $S(t_0) == S(t_1)$ is an initial sequent of PA$_\omega$ since $t_0 \approx t1$.\\
\\
\AxiomC{$S(t_0) == S(t_1)$}
\LeftLabel{\scriptsize$0,1$}
\RightLabel{Weakening}
\UnaryInfC{$\neg (t_0 == t_1) \vee (S(t_0) == S(t_1))$}
\DisplayProof
\\
\\
d) Derivation of Equality Implying Successor Equality when $t_0 \not\approx t1$\vspace*{-2mm}\\
\\
We know $\neg (t_0 == t_1)$ is an initial sequent of PA$_\omega$ since $t_0 \not\approx t1$.\\
\\
\AxiomC{$\neg (t_0 == t_1)$}
\LeftLabel{\scriptsize$0,1$}
\RightLabel{Weakening}
\UnaryInfC{$(S(t_0) == S(t_1)) \vee \neg (t_0 == t_1)$}
\LeftLabel{\scriptsize$0,1$}
\RightLabel{$E_1$}
\UnaryInfC{$\neg (t_0 == t_1) \vee (S(t_0) == S(t_1))$}
\DisplayProof
\\
\\
e) Derivation of Successor Equality Implying Equality when $S(t_0) \approx S(t1)$\vspace*{-2mm}\\
\\
We know $S(t_0) == S(t_1)$ is an axiom of PA$_\omega$ since $S(t_0) \approx S(t1)$.\\
\\
\AxiomC{$S(t_0) == S(t_1)$}
\LeftLabel{\scriptsize$0,1$}
\RightLabel{Weakening}
\UnaryInfC{$\neg (t_0 == t_1) \vee (S(t_0) == S(t_1))$}
\DisplayProof
\\
\\
f) Derivation of Successor Equality Implying Equality when $S(t_0) \not\approx S(t1)$\vspace*{-2mm}\\
\\
We know that $\neg (t_0 == t_1)$ is an initial sequent of PA$_\omega$ since $S(t_0) \not\approx S(t1)$.\\
\\
\AxiomC{$\neg (t_0 == t_1)$}
\LeftLabel{\scriptsize$0,1$}
\RightLabel{Weakening}
\UnaryInfC{$(S(t_0) == S(t_1)) \vee \neg (t_0 == t_1)$}
\LeftLabel{\scriptsize$0,1$}
\RightLabel{$E_1$}
\UnaryInfC{$\neg (t_0 == t_1) \vee (S(t_0) == S(t_1))$}
\DisplayProof
\\
\\
\end{tabular}
\end{example}

The generalised forms of the 5 remaining axioms are themselves initial sequents of PA$_\omega$, so the derivations are immediate.

\

\

The final element which needs to be reproduced is the axiom schema for induction.

\

\begin{definition}
The axiom schema for induction says that for any formula $A$ and free variable $x_n$, that the formula $A[x_n/0] \to (\forall t, A[x_n/t] \to A[x_n/S(t)]) \to \forall x, A[x_n/x]$ is a theorem of Peano Arithmetic.

The relevant part of the Coq code from the definition of the Calculus for Peano Arithmetic is repeated below.
\end{definition}

\begin{lstlisting}[language=Coq,label=list:indcution_defCoq,abovecaptionskip=-\medskipamount]
|induct : forall (A : formula) (n : nat),
            Peano_Theorem
              (lor (neg (substitution A n zero))
                  (lor (neg (univ n (lor (neg A) (substitution A n (succ (var n)))))
                      (univ n A))).
\end{lstlisting}

Since the formula for induction is universally quantified, it will necessarily involve the use of the $\omega$-rule. As such, it will be necessary to construct a version of $A[x_n/c]$ for every closed term $c$. The method for doing this will be to first construct something called the inductive iterate. 

\begin{definition}
  The inductive iterate $m$ of a formula $A$ is as follows:
\begin{center}
$I(m) := A[x_n/m] \vee \neg A[x_n/0] \vee \neg (\forall t, (\neg A[x_n/t] \vee A[x_n/S(t)]))$ for some $m \in \mathbb{N}$.

Or in normal notation.

$I(m) := A(m) \vee (A(0) \to \neg (\forall t, A(t) \to A(t+1)))$.
\end{center}

It is the statement: $A$ is true for $m$ or inductive reasoning does not hold for $A$.
\end{definition}

Since every closed term $c$ evaluates to some natural number $m$, assuming that we can derive $I(m)$ for all natural numbers, we can then use \lstinline{cut_cad} with a derivation of $\neg A[x_n/m] \vee A[x_n/c]$ to gain our set of premises for the $\omega$-rule. The derivation of $\neg A[x_n/m] \vee A[x_n/c]$ is again done using Theorem \ref{theorem:ELEM}.

\begin{definition}[Using LEM\_term to prepare for the $\omega$-rule]
\label{definition:omega_prep}
\normalfont

\

\scalebox{0.7}{\AxiomC{\hspace*{0.5cm}$\vdots$}

\AxiomC{$\vdots$}

\RightLabel{$I(m)$}
\UnaryInfC{$(A[x_i/m] \vee \neg A[x_i/0]) \vee \neg (\forall t, \neg A[x_i/t] \vee A[x_i/S(t)])$}
\RightLabel{$A_1E_1$}
\UnaryInfC{$(\neg A[x_i/0] \vee \neg (\forall t, \neg A[x_i/t] \vee A[x_i/S(t)])) \vee A[x_i/m]$}

\AxiomC{$\vdots$}
\RightLabel{LEM\_term $(A,i,m,c)$}
\UnaryInfC{$\neg A[x_i/m] \vee A[x_i/c]$}
\RightLabel{Cut}
\BinaryInfC{$(\neg A[x_i/0] \vee \neg (\forall t, \neg A[x_i/t] \vee A[x_i/S(t)])) \vee A[x_i/c]$}
\RightLabel{$E_1$}
\UnaryInfC{$A[x_i/c] \vee (\neg A[x_i/0] \vee \neg (\forall t, \neg A[x_i/t] \vee A[x_i/S(t)]))$}

\AxiomC{\hspace*{-2.5cm}$\vdots$\hspace*{0.5cm}}
\RightLabel{$\omega$-rule}
\TrinaryInfC{$(\forall t, A[x_i/t]) \vee (\neg A[x_i/0] \vee \neg (\forall t, \neg A[x_i/t] \vee A[x_i/S(t)]))$}
\RightLabel{$E_1A_1$}
\UnaryInfC{$\neg A[x_i/0] \vee (\neg (\forall t, \neg A[x_i/t] \vee A[x_i/S(t)]) \vee (\forall t, A[x_i/t]))$}
\DisplayProof}

\

\

\end{definition}

Therefore, deriving the induction schema instance for $A$ can be reduced to instead producing a derivation for all the inductive iterates of $A$, such that the derivations all have the same degree, and a height which is bounded by $m$ in some way, so as to ensure the validity of applying the $\omega$-rule. Building the inductive iterates will require an another intermediary, a structure called the inductive chain, which represents the idea of chaining together the inductive step some number of times.

\begin{definition}
The inductive chain of length $m$ of a formula $A$ is denoted $C(m)$ and is defined inductively.

$C(0)$ : The length 0 chain is the formula $\neg(\neg A[x_i/0] \vee A[x_i/1])$

$C(m+1)$ : The length $m+1$ chain is the formula $C(m) \vee \neg(\neg A[x_i/m+1] \vee A[x_i/m+2])$
\end{definition}
\begin{lstlisting}[language=Coq,label=list:indcutive_chain,abovecaptionskip=-\medskipamount]
Fixpoint inductive_chain (A : formula) (n m : nat) : formula :=
match m with
| 0 =>
    neg (lor (neg (substitution A n (represent 0))) 
             (substitution A n (succ (represent 0))))
| (S m') =>
    (lor (inductive_chain A n m')
         (neg (lor (neg (substitution A n (represent (S m')))) 
                   (substitution A n (succ (represent (S m')))))))
end.
\end{lstlisting}

The inductive chain is useful since every disjunct, once quantification has been applied, will be of the form $\neg (\forall t, A[x_n/t] \to A[x_n/S(t)]$), so after applying $m$ quantifications and contractions, (and numerous exchanges), $C(m)$ can be reduced to the final disjunct of $I(m)$. As such, if we can show that $A[x_n/m+1] \vee \neg A[x_n/0] \vee C(m)$ is derivable for all values of $m$, this would give us a derivation for all non-zero $I(m)$. These derivations can be found through induction over the value $m$. The base case is built from two instances of the Law of the Excluded Middle, which are then negated and weakened so they may be combined using the DeMorgan rule.

\begin{example}[Derivation for the Base Case]

\normalfont

\

\scalebox{0.7}{
\AxiomC{$\vdots$}
\RightLabel{LEM $A[x_i/0]$}
\UnaryInfC{$\neg A[x_i/0] \vee A[x_i/0]$}
\RightLabel{$E_1$}
\UnaryInfC{$A[x_i/0] \vee \neg A[x_i/0]$}
\RightLabel{Weakening}
\UnaryInfC{$A[x_i/1] \vee (A[x_i/0] \vee \neg A[x_i/0])$}
\RightLabel{$E_{1,2}A_2$}
\UnaryInfC{$A[x_i/0] \vee (\neg A[x_i/1] \vee A[x_i/0])$}
\RightLabel{Negation}
\UnaryInfC{$\neg\neg A[x_i/0] \vee (A[x_i/1] \vee \neg A[x_i/0])$}

\AxiomC{$\vdots$}
\RightLabel{LEM $A[x_i/1]$}
\UnaryInfC{$\neg A[x_i/1] \vee A[x_i/1]$}
\RightLabel{Weakening}
\UnaryInfC{$\neg A[x_i/0] \vee (\neg A[x_i/1] \vee A[x_i/1])$}
\RightLabel{$E_1A_1$}
\UnaryInfC{$\neg A[x_i/1] \vee (A[x_i/1] \vee \neg A[x_i/0])$}
\RightLabel{DeMorgan}
\BinaryInfC{$\neg (\neg A[x_i/0] \vee  A[x_i/1]) \vee (A[x_i/1] \vee \neg A[x_i/0])$}
\RightLabel{$A_2$}
\UnaryInfC{$(\neg (\neg A[x_i/0] \vee  A[x_i/1]) \vee A[x_i/1]) \vee \neg A[x_i/0]$}
\DisplayProof}
\end{example}

\

For the inductive case, we replace the instance of LEM on the left with the derivation from the inductive hypothesis and perform the same process. 

\

\begin{example}[Derivation of the Inductive Case]

\normalfont

\

\hspace*{-8mm}\scalebox{0.63}{
\AxiomC{$\vdots$}
\RightLabel{Inductive Hypothesis}
\UnaryInfC{$(C(m) \vee A[x_i/m+1]) \vee \neg A[x_i/0]$}
\RightLabel{Weakening}
\UnaryInfC{$A[x_i/m+2] \vee ((C(m) \vee A[x_i/m+1]) \vee \neg A[x_i/0])$}
\RightLabel{$E_{1,2,4,2,1}$}
\UnaryInfC{$A[x_i/m+1] \vee ((C(m) \vee A[x_i/m+2]) \vee \neg A[x_i/0])$}
\RightLabel{Negation}
\UnaryInfC{$\neg\neg A[x_i/m+1] \vee ((C(m) \vee A[x_i/m+2]) \vee \neg A[x_i/0])$}

\AxiomC{$\vdots$}
\RightLabel{LEM $(A[x_i/m+2])$}
\UnaryInfC{$\neg A[x_i/m+2] \vee A[x_i/m+2]$}
\RightLabel{Weakening}
\UnaryInfC{$(C(m) \vee \neg A[x_i/0]) \vee (\neg A[x_i/m+2] \vee A[x_i/m+2])$}
\RightLabel{$A_2E_{2,4,1}$}
\UnaryInfC{$\neg A[x_i/m+2] \vee ((C(m) \vee A[x_i/m+2]) \vee \neg A[x_i/0])$}

\RightLabel{DeMorgan}
\BinaryInfC{$\neg (\neg A[x_i/m+1] \vee  A[x_i/m+2])\vee ((C(m) \vee A[x_i/m+2]) \vee \neg A[x_i/0])$}
\RightLabel{$E_{1,2,4}$}
\UnaryInfC{$((C(m)\vee \neg (\neg A[x_i/m+1] \vee  A[x_i/m+2])) \vee A[x_i/m+2]) \vee \neg A[x_i/0]$}
\RightLabel{Definition of $C$}
\UnaryInfC{$(C(m+1) \vee A[x_i/m+2]) \vee \neg A[x_i/0]$}
\DisplayProof}

\end{example}

\

With this induction complete we have completed derivations of $I(m)$ for all non-zero $m$. Finally $I(0)$ is just a weakening of the Law of the Exclude Middle applied to $A[x_n/0]$.

\

\begin{example}[Derivation of the Initial Inductive Iterate]

\

\normalfont 

\AxiomC{$\vdots$}
\RightLabel{LEM $A[x_i/0]$}
\UnaryInfC{$\neg A[x_i/0] \vee A[x_i/0]$}
\RightLabel{$E_1$}
\UnaryInfC{$A[x_i/0] \vee \neg A[x_i/0]$}
\RightLabel{Weakening}
\UnaryInfC{$\neg (\forall t, \neg A[x_i/t] \vee A[x_i/S(t)]) \vee (A[x_i/0] \vee \neg A[x_i/0])$}
\RightLabel{$E_1$}
\UnaryInfC{$(A[x_i/0] \vee \neg A[x_i/0]) \vee \neg (\forall t, \neg A[x_i/t] \vee A[x_i/S(t)])$}
\DisplayProof
\end{example}

\

\

Now with all the inductive iterates constructed, we are almost ready to apply the process outlined in Definition \ref{definition:omega_prep} to them to complete the derivation of the induction schema. The only thing left to do it check the uniformity of degree and the bound on height. To simplify the diagrams we have been omitting the decorations from the sequents, however it is very important to keep track of these when the ultimate goal to to apply an omega rule. If you recall from our definition of LEM and LEM\_term, these derivations where all of degree 0 and had a height which depends only on the number of logical connectives in the formula being instantiated. As such, every instance of LEM from Definition \ref{definition:omega_prep} onwards has been of the same height (which we shall call $\alpha$). That definition was also the only time cut was used in this whole section, so we know the degree of all the inductive iterates is 0. All of the cuts in Definition \ref{definition:omega_prep} are on formulae of the form $A[x_i/m]$ for varying values of $m$, and as such those cuts all have the same degree, satisfying that requirement. We can now use $\alpha$ to determine what the heights of all the inductive iterates are and find a bound.

\

The derivation of $I(0)$ only used a single tall rule after the LEM instance so it has height $\alpha + 1$, and the longest path in the derivation of $I(m)$ uses $4$ tall rules plus the height of the previous derivation, giving a height of $\alpha + 4m + 1$. As such, since $\alpha$ itself was also finite, all of the premises for the application of the $\omega$-rule in Definition \ref{definition:omega_prep} are finite, so are bounded by $\omega$ and the derivation shown in the deductive rule has its side condition satisfied, completing our derivation of the Induction Schema.

\

Therefore, as PA$_\omega$ admits both deductive rules of Peano Arithmetic, and the closure of every theorem in Peano Arithmetic is a theorem in PA$_\omega$ it is clear that the consistency result for PA$_\omega$ must now extend to Peano Arithmetic. Since if Peano Arithmetic was inconsistent, both the closure of $A$ and $\neg A$ would have to be theorems in PA$_\omega$, which we know is not true. This completes the proof of the consistency of Peano Arithmetic.

\begin{theorem}[Peano Arithmetic is consistent]
If a formula $A$ is a theorem of Peano Arithmetic, then $\neg A$ is not a theorem of Peano Arithmetic.\vspace*{-3mm} 
\end{theorem}
\begin{lstlisting}[language=Coq,label=list:PA_con_Coq,abovecaptionskip=-\medskipamount]
Theorem PA_Consistent :
    forall (A : formula) (n1 n2 : nat) (alpha1 alpha2 : ord),
        Peano_Theorem_Base A n1 alpha1 ->
            Peano_Theorem_Base (neg A) n2 alpha2 -> False.
Proof.
intros.
pose proof(PA_Base_closed_PA_omega _ _ _ H (repr 0) (repr_closed _)).
pose proof(PA_Base_closed_PA_omega _ _ _ H0 (repr 0) (repr_closed _)).
rewrite closure_neg in H2; auto.
apply (inconsistent_danger _ _ _ _ _ H1 H2). 
Qed.
\end{lstlisting}

\section{Related Works}

\subsection{Finitely Represented Ordinals In Coq}

The definition of finitely representable ordinals, their ordering and normal form, along with most of their arithmetic properties and the definition of strong induction, were all taken from existing work by Pierre Cast\`{e}ran and Evelyne Contejean\cite{casteran}. During our earlier efforts, we only used the main definition, but we found ourselves unable to prove strong induction held, even over the normal-form fragment. As such, we ended up importing larges portions of that library to use as a dependencies for our code, so that we could use those results which they had already proved. This was a great boon to our work, as it allowed us to instead focus on the aspects of the ordinals which were specific to the problem at hand, such as monotonicity results; the existence of certain inequalities and decomposition's of limit ordinals. 

\subsection{Other Attempts at Formalising Gentzen's Results}

The initial inspiration for this work was a masters thesis by Morgan Sinclaire\cite{sinclaire}, which was an incomplete attempt at formalising Gentzen's proof in Coq as outlined in the book ``Introduction to Mathematical Logic'' by Elliott Mendelson\cite{mendelson}. Sinclaire had seemingly abandoned the work and was not contactable, so we took the framework that he had provided and completed it, along with fixing some of the existing bugs. Work that had already been done by Sinclaire include: the definition of the language for PA$_\omega$ and facts about substitution and free variables; a definition of finitely representable ordinals inspired by the work above, some of the machinery associated with the ordinals that would be needed for the later parts of the proof; a definition for theorems of PA$_\omega$ without decorations; the proof that the Law of the Excluded Middle was on of those theorems; the definition of proof trees for PA$_\omega$ with decorations; and inversion principles for the Negation and DeMorgan rules. Notable things that had yet to be completed included, the necessary inequalities over the ordinals, strong induction over the ordinals, an inversion for the $\omega$-rule, the cut elimination argument itself\footnote{Although there was a series of theorems without proof defined to give an outline of the intended argument}, and the definition of the Hilbert Calculus for Peano Arithmetic and its equivalence to PA$_\omega$. Bugs that needed to be fixed included: the definition of ordinal exponentiation, which was incorrect and had no fixed points; the lack of decorations on the theorems of PA$_\omega$, which are necessary to show that all proof trees can only have theorems as endsequents; the assignment of heights to proof trees, since several of the tall rules did not increase the height; and all proofs which became invalidated by any of the previous changes, which included all the results relating to the Law of the Excluded Middle and inverting any of the deductive rules.

\begin{appendices}

\section{Derivations of FOL Axioms in PA$_\omega$}
\label{secA1}

Derivation for FOL2 : $(A \to (B \to C)) \to ((A \to B) \to (A \to C))$

\scalebox{0.7}{

\AxiomC{$\vdots$}
\RightLabel{LEM $(\neg A \vee B)$}
\UnaryInfC{$\neg (\neg A \vee B) \vee (\neg A \vee B)$}
\RightLabel{$E_1,E_2$}
\UnaryInfC{$(\neg A \vee \neg (\neg A \vee B)) \vee B$}

\AxiomC{$\vdots$}
\RightLabel{LEM $(\neg A \vee (\neg B \vee C))$}
\UnaryInfC{$\neg (\neg A \vee (\neg B \vee C)) \vee (\neg A \vee (\neg B \vee C))$}
\RightLabel{$A_2E_{1,3}A_2E_1A_2E_{3,4,2,4,1}$}
\UnaryInfC{$\neg B \vee ((C \vee \neg (\neg A \vee (\neg B \vee C))) \vee \neg A)$}

\RightLabel{Cut $(B)$}
\BinaryInfC{$(\neg A \vee \neg (\neg A \vee B)) \vee ((C \vee \neg (\neg A \vee (\neg B \vee C))) \vee \neg A)$}
\RightLabel{$A_2E_{2,4}A_1$}
\UnaryInfC{$((\neg A) \vee (\neg A)) \vee ((\neg ((\neg A) \vee B)) \vee (C \vee (\neg ((\neg A) \vee ((\neg B) \vee C)))))$}
\RightLabel{Contraction}
\UnaryInfC{$(\neg A) \vee ((\neg ((\neg A) \vee B)) \vee (C \vee (\neg ((\neg A) \vee ((\neg B) \vee C)))))$}
\RightLabel{$A_{2,2}E_{4,3,2,3,1,3,2,3,2}A_{1,1}$}
\UnaryInfC{$(\neg ((\neg A) \vee ((\neg B) \vee C))) \vee (\neg ((\neg A) \vee B) \vee ((\neg A) \vee C))$}
\DisplayProof
}

\

\

Derivation of FOL3 : $ (\neg A \to \neg B) \to ((\neg A \to B) \to A)$

\scalebox{0.7}{
\AxiomC{$\vdots$}
\RightLabel{LEM ($A$)}
\UnaryInfC{$\neg A \vee A$}
\RightLabel{Negation}
\UnaryInfC{$\neg\neg\neg A \vee A$}
\RightLabel{Weakening}
\UnaryInfC{$\neg (\neg\neg A \vee B) \vee (\neg\neg\neg A \vee A)$}
\RightLabel{$E_{1,2}A_1$}
\UnaryInfC{$\neg\neg\neg A \vee (\neg (\neg\neg A \vee B) \vee A)$}

\AxiomC{$\vdots$}
\RightLabel{LEM ($A$)}
\UnaryInfC{$\neg A \vee A$}
\RightLabel{Weakening}
\UnaryInfC{$\neg\neg B \vee (\neg A \vee A)$}
\RightLabel{$E_1A_1$}
\UnaryInfC{$\neg A \vee (A \vee \neg\neg B)$}
\RightLabel{Negation}
\UnaryInfC{$\neg\neg\neg A \vee (A \vee \neg\neg B)$}

\AxiomC{$\vdots$}
\RightLabel{LEM ($\neg B$)}
\UnaryInfC{$\neg\neg B \vee \neg B$}
\RightLabel{$E_1$}
\UnaryInfC{$\neg B \vee \neg\neg B$}
\RightLabel{Weakening}
\UnaryInfC{$A \vee (\neg B \vee \neg\neg B)$}
\RightLabel{$A_2E_3A_1$}
\UnaryInfC{$\neg B \vee (A \vee \neg\neg B)$}

\RightLabel{DeMorgan}
\BinaryInfC{$\neg (\neg\neg A \vee B) \vee (A \vee \neg\neg B)$}
\RightLabel{$A_2E_1$}
\UnaryInfC{$\neg\neg B \vee (\neg (\neg\neg A \vee B) \vee A)$}

\RightLabel{DeMorgan}
\BinaryInfC{$\neg(\neg\neg A \vee \neg B) \vee (\neg (\neg\neg A \vee B) \vee A)$}
\DisplayProof
}

\

\

Derivation of FOL4 : $\forall x, A(x) \to A(m)$ for any $m \in \mathbb{N}$.

\scalebox{0.7}{
\AxiomC{$\vdots$}
\RightLabel{LEM $(A[x_i/m]$}
\UnaryInfC{$\neg A[x_i/m] \vee A[x_i/m]$}
\RightLabel{Quantification}
\UnaryInfC{$\neg(\forall t, A[x_i/t]) \vee A{x_i/m} $}
\DisplayProof}

\

\

Derivation of FOL5 : $\forall x, (A \vee B)(x) -> A \vee (\forall x, B(x))$ if $A$ has no free variables.

\scalebox{0.7}{
\AxiomC{\hspace{4mm}$\vdots$}

\AxiomC{$\vdots$}
\RightLabel{LEM $(\neg A \vee B[x_i/m])$}
\UnaryInfC{$\neg(\neg A \vee B[x_i/m]) \vee (\neg A \vee B[x_i/m])$}
\RightLabel{****}
\UnaryInfC{$\neg(\neg A \vee B)[x_i/m] \vee (\neg A \vee B[x_i/m])$}
\RightLabel{Quantification}
\UnaryInfC{$\neg (\forall t, (\neg A \vee B)[x_i/t]) \vee (\neg A \vee B[x_i/m])$}
\RightLabel{$A_2E_1$}
\UnaryInfC{$B[x_i/m] \vee (\neg (\forall t, (\neg A \vee B)[x_i/t]) \vee \neg A)$}
\AxiomC{$\vdots$\hspace*{4mm}}
\RightLabel{$\omega$-rule}
\TrinaryInfC{$(\forall t, B[x_i/t]) \vee (\neg (\forall t, (\neg A \vee B)[x_i/t]) \vee \neg A)$}
\RightLabel{$E_1A_1$}
\UnaryInfC{$\neg (\forall t, (\neg A \vee B)[x_i/t]) \vee (\neg A \vee (\forall t, B[x_i/t]))$}
\DisplayProof}

\

Where the line marked $****$ is not a logical deduction, but an equality of formulae which holds since $x_i$ was assumed to not be free in $A$, meaning the substitution is trivial.




\end{appendices}


\bibliography{sn-bibliography}% common bib file

\end{document}